\documentclass[journal]{IEEEtran}
\newcommand{\figref}{Fig.~\ref}

\usepackage{cite,graphicx,amsmath,psfrag,bm}
\usepackage{subfigure}
\usepackage[usenames,dvipsnames]{xcolor}
\usepackage{mathabx}
\usepackage{cancel}
\usepackage{epstopdf}
\usepackage{pstool}
\usepackage{caption}

\begin{document}

\title{A High-Resolution Transmission-Type (TT) Phaser Based on Reflection-Type (RT) Units for Radio Analog Signal Processing (R-ASP)}

\author{Lianfeng Zou,~\IEEEmembership{Student member,~IEEE,}
        Christophe Caloz,~\IEEEmembership{Fellow,~IEEE}
\thanks{L. Zou and C. Caloz are with the Department
of Electrical Engineering and Poly-Grames research center, \'{E}cole Polytechnique de Montr\'{e}al, Montr\'{e}al, Qu\'{e}bec, H3T 1J4, Canada. e-mail: lianfeng.zou@polymtl.ca.}%
}

\maketitle

\begin{abstract}
A high Radio Analog Signal Processing (R-ASP) resolution transmission-type (TT) phaser based on reflection-type (RT) phaser units is introduced, theoretically studied and experimentally demonstrated. It is first shown that RT phasers inherently exhibit higher R-ASP resolution than their TT counterparts because their group delay swing is proportional to the reflection coefficient associated with a resonator coupling mechanism (admittance inverter), easy to maximize towards unity, rather than to a coupled-line coupling coefficient, typically restricted to values will inferior to unity, as in the RT case. Moreover, a detailed sensitivity analysis reveals that the proposed phaser is simultaneously features high R-ASP resolution and low sensitivity to fabrication tolerance, which makes it an ideal solution for R-ASP. The proposed phaser exhibits a 5~ns group delay swing over a fractional bandwidth of about 50$\%$ around 4~GHz. 
\end{abstract}

\begin{IEEEkeywords}
Radio Analog Signal Processing (R-ASP), phasers, dispersion engineering, coupled-line couplers, sensitivity analysis.
\end{IEEEkeywords}

\IEEEpeerreviewmaketitle

\section{Introduction}\label{SEC:INTRO}

\IEEEPARstart{R}{adio} analog signal processing (R-ASP)~\cite{Jour:2013_MwMag_Caloz} is new microwave, millimeter-wave and térahertz technology, inspired by ultra-fast optics~\cite{BK:2007_Saleh,JOUR:2010_Azana}, that consists in processing radio signals in their pristine analog form, using highly dispersive components called ``phasers'' for faster, lower-cost and frequency-scalable RF. This technology is still in its infancy, but several R-ASP applications have already been reported, including a compressive receiver based on CRLH dispersive meta-lines~\cite{JOUR:2009_TMTT_Abielmona}, a CRLH-based PPM transmitter~\cite{JOUR:2008_MWCL_Nguyen}, a real-time spectrum analyzer~\cite{JOUR:2003_TMTT_Laso}, a real-time spectrum sniffer~\cite{JOUR:2012_MWCL_Nikfal}, time-stretching system for enhanced sampling~\cite{JOUR:2007_TMTT_Schwartz,JOUR:2012_TMTT_Xiang}, a time-compressor for increased data rate~\cite{CONF:2008_IRWS_Schwartz}, a real-time spectrogram analyzer~\cite{JOUR:2009_TMTT_Gupta} and dispersion-coded RFIDs~\cite{JOUR:2011_AWPL_Gupta}.

Phasers, which are components exhibiting specified group delay versus frequency responses to accomplish various communication or instrumentation applications, are the key processing units in R-ASP~\cite{Jour:2013_MwMag_Caloz}. Their figure of merit is the real-time frequency resolution, which is proportional to their group delay swing, $\Delta\tau$. They may be implemented in various technologies, with various benefits and drawbacks: 1)~Edge-coupled C-section phasers~\cite{JOUR:2010_TMTT_Gupta, JOUR:1966_TMTT_Cristal}, which are easy to design but limited in $\Delta\tau$ by practically achievable coupling levels~\cite{BK:2007_Mongia}, 2)~broad-side coupled C-section~\cite{JOUR:2012_MWCL_Horii}, which provide higher $\Delta\tau$ but require vias and multilayered configurations, 3)~CRLH phasers~\cite{JOUR:2011_PROC_Caloz,JOUR:2012_TMTT_Shulabh}, which provide high $\Delta\tau$ in uniplanar configurations but involve significant design complexity, 4)~Bragg grating phasers~\cite{JOUR:2003_TMTT_Laso}, which may provide very high $\Delta\tau$ but suffer from delay ripples due to multiple reflections, and 4)~reflection-type coupled-resonator based phasers~\cite{JOUR:2012_TMTT_Zhang}, which may also provide high $\Delta\tau$ but suffer from narrow bandwidth.

In~\cite{JOUR:2013_EL_Zhang}, it was shown that reflection-type (RT) phasers (one-port networks) inherently provide higher group delay swing than transmission-type (TT) phasers (two-port networks). Based on this observation, a combined phaser formed by combining pairs of RT phasers through a hybrid coupler was proposed to leverage the R-ASP performance of RT phasers with two-port advantage of TT phasers was proposed in the conference paper~\cite{CONF:2014_IMS_Zou} associated with the present special issue. The present paper expands on this work, providing in addition a)~analytical group delay swing formulas that emphasize the fundamental differences between TT and RT phasers (Sec.~\ref{SEC:COMPARISON_TT_RT}), b)~a scattering description of the combined phaser allowing to separate the effects of the RT phaser units and the hybrid coupler (Sec.~\ref{SEC:CONCEPT_TR_PHASER}), c)~a complete analysis of the coupled-line RT phaser units (Sec.~\ref{SEC:ANALYSIS}), d)~a change of the branch-line hybrid coupler to a much broader bandwidth slot coupled-line hybrid coupler and showing that this coupler is essentially transparent to R-ASP (Sec.~\ref{SEC:HYBRID}), e)~a simple synthesis technique based on analytical formulas developed in c) (Sec.~\ref{SEC:SYNTHESIS}), f)~an experimental demonstration of a 6-unit combined phaser based on the synthesis performed in e) with a group delay swing of 5~ns over a fractional bandwith of $50\%$ (Sec.~\ref{SEC:IMP}), g)~a detailed sensitivity analysis showing that the combined phaser features minimal relative sensitivity to tolerance in the range where it provides maximal R-ASP resolution (Sec.~\ref{SEC:SENSITIVITY}).

\section{Comparison of Transmission-Type (TT) and Reflection-Type (RT) Phaser}\label{SEC:COMPARISON_TT_RT}

\begin{figure}[h!t]
   \centering
   \psfragfig*[width=0.95\linewidth, trim={0in 0in 0in 0in}]{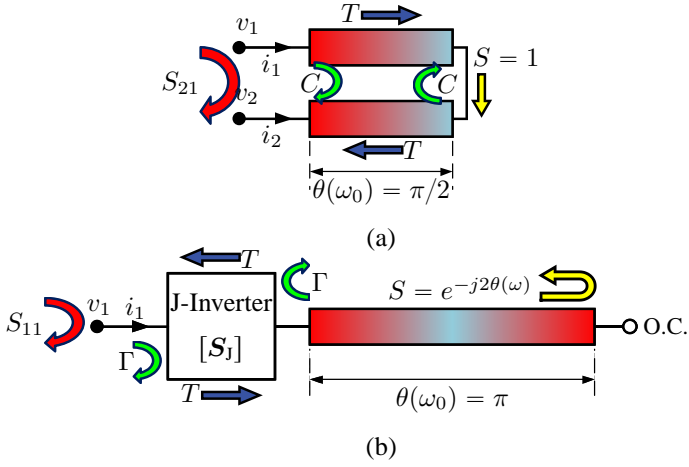}{
    \psfrag{A}[c][c][1]{$v_1$}
    \psfrag{X}[c][c][1]{$i_1$}
    \psfrag{B}[c][c][1]{$v_2$}
    \psfrag{Y}[c][c][1]{$i_2$}
    \psfrag{C}[c][c][1]{$C$}
    \psfrag{Z}[l][l][0.9]{O.C.}
    \psfrag{k}[c][c][1]{$\Gamma$}
    \psfrag{E}[c][c][1]{$\theta(\omega_0)=\pi/2$}
    \psfrag{F}[c][c][1]{$\theta(\omega_0)=\pi$}
    \psfrag{a}[c][r][1]{J-Inverter}
    \psfrag{b}[c][r][1]{$\boldsymbol{[S_{\text{J}}]}$ }
    \psfrag{g}[c][c][1]{(a)}
    \psfrag{h}[c][c][1]{(b)}
    \psfrag{U}[r][r][1]{$S_{21}$}
    \psfrag{V}[r][r][1]{$S_{11}$}
     \psfrag{T}[c][c][1]{$T$}
     \psfrag{S}[l][l][1]{$S=1$}
     \psfrag{R}[r][r][1]{$S=e^{-j2\theta(\omega)}$}
  }
   \caption{Two main types of phasers. (a) Transmission-type (TT) phaser. (b) Reflection-type (RT) phaser.}
   \label{FIG:IDEAL_UNITS}
\end{figure}

There are two main types of phasers, namely transmission-type (TT) and reflection-type (RT) phasers, shown in Figs.~\ref{FIG:IDEAL_UNITS}(a) and~\ref{FIG:IDEAL_UNITS}(a), respectively. A TT phaser is formed by interconnecting the two end ports of a quarter-wavelength coupled-line coupler and the resulting two-port structure is called a ``C-section''. An RT phaser is one-port network formed by a half-wavelength open-ended resonator coupled to the port through an admittance inverter. The responses of TT and RT phasers were compared in terms of the coupling factor and inversion factor, respectively, in~\cite{JOUR:2013_EL_Zhang}. This section characterizes the RT phaser in terms of the reflection coefficient (instead of the inversion factor), which provides more insightful comparison, and also  defines the key quantities and notations for the remaining of the paper.

The coupled and through responses of a quarter-wavelength coupled-line coupler  are~\cite{BK:2007_Mongia}
\begin{subequations}\label{EQ:RESP_BWD}
\begin{equation}\label{EQ:RESP_BWD_CPL_}
  C=\dfrac{jk\sin\theta}{\sqrt{1-k^2}\cos\theta+j\sin\theta},
\end{equation}
and
\begin{equation}\label{EQ:RESP_BWD_TRH_}
  T =\dfrac{\sqrt{1-k^2}}{\sqrt{1-k^2}\cos\theta+j\sin\theta},
\end{equation}
\end{subequations}
respectively, where $\theta$ is the electrical length of the structure and $k=C(\theta=\pi/2)$ is the maximal backward coupling coefficient. Defining $\omega_0$ as the frequency where the coupler exhibits maximum coupling, i.e. when it is a quarter-wavelength (or $\pi/2$ in terms of phase) long, we have $\theta=\pi\omega/(2\omega_0)$, and $k=C(\omega_0)$.

The transfer (transmission) function of the TT phaser [Fig.~\ref{FIG:IDEAL_UNITS}(a)], assuming matching ($S_{11}^{\text{T}} = 0$),  may be constructed from multiple \emph{coupling} events as
\begin{equation}\label{EQ:RESP_CSEC_INTERF}
\begin{split}
  S_{21}^{\text{T}}(\omega) &= C+TST+TSCST+TSCSCST+\ldots\\
  &= C+T^2\sum_{n=0}^{\infty}C^nS^{n+1}\\
  &= C+\dfrac{ST^2}{1-SC},
\end{split}
\end{equation}
where $S$ is the the the transfer function of the interconnection at the end of the coupler. Inserting \eqref{EQ:RESP_BWD} into \eqref{EQ:RESP_CSEC_INTERF}, and assuming that the interconnection is a simple short ($S=1$) yields
\begin{subequations}\label{EQ:RESP_CSEC}
\begin{equation}\label{EQ:RESP_CSEC_S21}
  S_{21}^{\text{T}}=\dfrac{1-j\rho_{\text{T}}\tan(\pi\omega/2\omega_0)}{1+j\rho_{\text{T}}\tan(\pi\omega/2\omega_0)},
\end{equation}
with
\begin{equation}\label{EQ:RESP_CSEC_PARAM}
 \rho_{\text{T}} = \sqrt{\dfrac{1-k}{1+k}},
\end{equation}
\end{subequations}
where $\rho_{\text{T}}$ is a parameter that will be useful in later comparison. It is immediately apparent in~\eqref{EQ:RESP_CSEC_S21} that a TT phaser is an all-pass network ($\left|S_{21}^{\text{T}}\right|=1$). The normalized group delay response with respect to the harmonic period at $\omega_0$, $T_0=1/f_0=2\pi/\omega_0$, is then obtained by taking the derivative of $\angle S_{21}^{\text{T}}(\omega)$ in~\eqref{EQ:RESP_CSEC_S21}. The result is
\begin{equation}\label{EQ:RESP_CSEC_GD}
\overline{\tau}^{\text{T}}(\omega)=\dfrac{\tau^{\text{T}}(\omega)}{T_0}=\dfrac{\rho_{\text{T}}/2}{(\rho_{\text{T}}^2-1)\sin^2(\pi\omega/2\omega_0)+1},
\end{equation}

and the normalized group delay swing is then
\begin{equation}\label{EQ:RESP_CSEC_GDSWG}
\begin{split}
    \Delta\overline{\tau}^{\text{T}}
    &=\overline{\tau}^{\text{T}}_{\text{max}}-\overline{\tau}^{\text{T}}_{\text{min}}\\
    &=\left.\overline{\tau}^{\text{T}}\right|_{\omega=(2n-1)\omega_0}
     -\left.\overline{\tau}^{\text{T}}\right|_{\omega=2n\omega_0}\\
    &=\dfrac{1}{2}\left(\dfrac{1}{\rho_{\text{T}}}-\rho_{\text{T}}\right)\\
    &=\dfrac{k}{\sqrt{(1+k)(1-k)}}.
\end{split}
\end{equation}

In the RT phaser [Fig.~\ref{FIG:IDEAL_UNITS}(b)], the coupling mechanism can be modeled by a J-inverter~\cite{BK:1980_Matthaei}, whose $ABCD$ matrix reads
\begin{equation}\label{EQ:JINV_DEF}
  \left[
  \begin{array}{cc}
    A & B\\[0.1in]
    C & D
  \end{array}
  \right]
  =
  \left[
  \begin{array}{cc}
    0 & \pm{j}{1/J}\\[0.1in]
    \pm{j}{J} & 0
  \end{array}
  \right].
\end{equation}
Using standard network conversion formulas~\cite{BK:2011_Pozar}, the corresponding matched-output reflection coefficient is found as
\begin{equation}\label{EQ:GammaR}
\Gamma=\frac{1-\left(JZ_0\right)^2}{1+\left(JZ_0\right)^2},
\end{equation}
which may be solved for $J$ to give
\begin{subequations}\label{EQ:JrhoR}
\begin{equation}\label{EQ:J_fc_rhoR}
J = \dfrac{1}{Z_0\sqrt{\rho_\text{R}}},
\end{equation}
where
\begin{equation}\label{EQ:JrhoR}
\rho_{\text{R}} = \dfrac{1+|\Gamma|}{1-|\Gamma|},
\end{equation}
\end{subequations}
where $\rho_{\text{R}}$ is identified as the standing wave ratio (SWR).

The transfer (reflection) function of the RT phaser may be constructed from multiple \emph{reflection} events procedure as
\begin{equation}\label{EQ:RESP_CPLRES_INTERF}
  \begin{split}
    S_{11}^{\text{R}}(\omega)&=\Gamma+TST+TS\Gamma ST+TS\Gamma S\Gamma ST\\
    &=\Gamma + T^2\sum_{n=0}^{\infty}\Gamma^nS^{n+1}\\
    &= \Gamma+\dfrac{ST^2}{1-S\Gamma },
  \end{split}
\end{equation}
where $S=e^{-j2\theta}$  is the round-trip (including the open-end) transfer function of the resonator, which is half-wavelength (or $\pi$ in terms of phase) at $\omega_0$, so that now $\theta=\pi\omega/\omega_0$, and where $T$ is the transmission coefficient through the inverter, which is a is assumed to be lossless, so that $|T|=\sqrt{1-|\Gamma|^2}$. Inserting then \eqref{EQ:GammaR} with \eqref{EQ:JrhoR} into \eqref{EQ:RESP_CPLRES_INTERF} yields
\begin{equation}\label{EQ:RESP_CPLRES_S11}
    S_{11}^{\text{R}}=\dfrac{1-j\rho_{\text{R}}\tan(\pi\omega/\omega_0)}{1+j\rho_{\text{R}}\tan(\pi\omega/\omega_0)},
\end{equation}
where $\rho_{\text{R}}$ was defined in \eqref{EQ:JrhoR}, and which has the form of an all-pass one-port network. It appears that~\eqref{EQ:RESP_CPLRES_S11} is formally similar to \eqref{EQ:RESP_CSEC_S21}, expect for the factor $1/2$ in the argument of the tangent. The normalized group delay response of the RT phaser is obtained by taking the derivative of $S_{11}^{\text{R}}$ in~\eqref{EQ:RESP_CPLRES_S11}. This results in
\begin{equation}\label{EQ:RESP_CPLRES_GD}
\overline{\tau}^{\text{R}}(\omega)=\dfrac{\tau^{\text{R}}(\omega)}{T_0}=\dfrac{\rho_{\text{R}}}{(\rho_{\text{R}}^2-1)\sin^2(\pi\omega/\omega_0)+1},
\end{equation}
which is also formally similar to the normalized group delay response of the TT phaser, Eq.~\eqref{EQ:RESP_CSEC_GD}. The corresponding normalized group delay swing is
\begin{equation}\label{EQ:RESP_CPLRES_GDSWG}
\begin{split}
    \Delta\overline{\tau}^{\text{R}}
    &=\overline{\tau}^{\text{R}}_{\text{max}}-\overline{\tau}^{\text{R}}_{\text{min}}\\
    &=\left.\overline{\tau}^{\text{R}}\right|_{\omega=n\omega_0}
     -\left.\overline{\tau}^{\text{R}}\right|_{\omega=(2n-1)\omega_0/2}\\
    &=\rho_{\text{R}}-\dfrac{1}{\rho_{\text{R}}}\\
    &=\dfrac{4|\Gamma|}{(1+|\Gamma|)(1-|\Gamma|)}.
\end{split}
\end{equation}

In summary, the RT and TT phasers shown in \figref{FIG:IDEAL_UNITS} are both all-pass systems and they share identical transfer function and group delay forms. However, the parameters $\rho_\text{T}$ and $\rho_\text{R}$ represent different physical mechanisms: $\Delta\overline{\tau}^{\text{T}}$ is (nonlinearily) proportional to the backward coupling factor ($k$) of the coupled-line coupler, whereas $\Delta\overline{\tau}^{\text{R}}$ is (nonlinearly) proportional to the magnitude of the reflection coefficient ($|\Gamma|$) of the admittance inverter. Since a high coupling coefficient is more difficult to achieve~\cite{BK:2007_Mongia} than a high reflection coefficient, the RT phaser practically provides higher group delay swings than its TT counterpart.

\section{Concept of Transmission-Type Phaser Based on Reflection-Type Units (TT-RT Phaser)}\label{SEC:CONCEPT_TR_PHASER}

It has been seen in the previous section that RT phasers are inherently more dispersive than TT phasers, meaning that they exhibit higher group delay swings and thus higher R-ASP resolution~\cite{Jour:2013_MwMag_Caloz}. However, the  signal generated by RT phasers cannot be straightforwardly separated from the input signal. Therefore, it makes sense to ask whether RT phaser units might be combined to form two-port network phasers for a novel type of phaser providing the benefits of both types of phasers without suffering from their respective drawbacks.

\begin{figure}[h!t]
\centering
  \psfragfig*[width=0.95\linewidth, trim={0in 0in 0in 0in}]{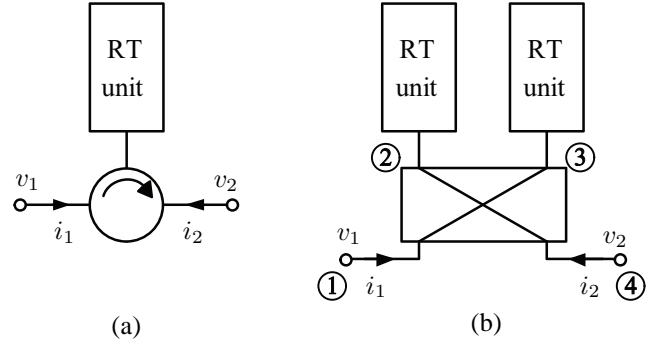}{
  \psfrag{a}[c][c][1]{(a)}
  \psfrag{b}[c][c][1]{(b)}
  \psfrag{e}[l][l][1]{$v_1$}
  \psfrag{f}[l][l][1]{$i_1$}
  \psfrag{g}[r][r][1]{$v_2$}
  \psfrag{h}[r][r][1]{$i_2$}
  }
  \caption{Two possible configurations for combining RT phasers into a TT phaser, using (a) a circulator, and (b) a hybrid coupler.}
  \label{FIG:TT_RT_UNIT}
\end{figure}

Several approaches might be envisioned for such a combination. The conceptually simplest of them would consist in using a circulator, as illustrated in \figref{FIG:IDEAL_UNITS}(a). However, a circular requires a permanent magnet, it is bulky and non-integrable, and, even worse, it exhibits a small bandwidth that would typically alter the bandwidth performance of the RT units. Therefore, this approach is not pursued here. A more elaborate and practical approach, chosen here, consists in resorting to a broadband hybrid coupler combining two phaser units as shown in~\figref{FIG:TT_RT_UNIT}(b). This configuration is similar to that of hybrid reflection-type phase shifters~\cite{JOUR:1995_TMTT_Lucyszyn}, except that the (reflection-type) phase shifters are replaced here by RT phasers

\begin{figure}[h!t]
\centering
  \psfragfig*[width=1\linewidth, trim={-0.05in 0in -0.4in 0in}]{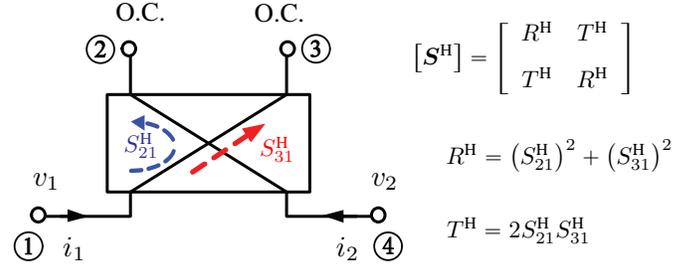}{
  \psfrag{e}[Tl][Bl][1.1]{$v_1$}
  \psfrag{f}[Bl][Tl][1.1]{$i_1$}
  \psfrag{g}[Tr][Br][1.1]{$v_2$}
  \psfrag{h}[Br][Tr][1.1]{$i_2$}
  \psfrag{o}[l][l][1]{O.C.}
  \psfrag{c}[r][r][1]{O.C.}
  \psfrag{t}[l][l][0.9]{\textcolor{red}{$S_{31}^\text{H}$}}
  \psfrag{k}[r][r][0.9]{\textcolor{Blue}{$S_{21}^\text{H}$}}
  \psfrag{R}[l][l][0.9]{$R^{\text{H}}=\left(S_{21}^\text{H}\right)^2+\left(S_{31}^\text{H}\right)^2$}
  \psfrag{T}[l][l][0.9]{$T^{\text{H}}=2S_{21}^\text{H}S_{31}^\text{H}$}
  \psfrag{s}[l][l][0.9]{$\boldsymbol{\left[S^{\text{H}}\right]}=\left[\begin{array}{cc}
    R^{\text{H}} & T^{\text{H}}\\[0.1in]
    T^{\text{H}} & R^{\text{H}}
  \end{array}\right]$ }
  }
  \caption{Hybrid coupler circuit with open terminations \mbox{($S_{11}^\text{R}=1$)} generating $S_{11}^\text{C}=R^\text{H}$ and $S_{21}^\text{C}=T^\text{H}$ in \eqref{EQ:RESP_TT_RT_UNIT}.}
  \label{FIG:HYB_2PORTS_OPEN}
\end{figure}

Assume the coupled and through responses of the hybrid coupler in \figref{FIG:TT_RT_UNIT}(b) are $S_{21}^\text{H}$ and  $S_{31}^\text{H}$, respectively. The scattering parameters of the phaser combining the two RT phaser units via the hybrid are obtained, with the help of~\figref{FIG:HYB_2PORTS_OPEN}, as
\begin{subequations}\label{EQ:RESP_TT_RT_UNIT}
\begin{equation}\label{EQ:RESP_TT_RT_UNIT_S11}
  S_{11}^{\text{C}} =
  \left[\left(S_{21}^\text{H}\right)^2+\left(S_{31}^\text{H}\right)^2\right]S_{11}^{\text{R}}=R^\text{H}S_{11}^{\text{R}},
\end{equation}
\begin{equation}\label{EQ:RESP_TT_RT_UNIT_S21}
  S_{21}^{\text{C}} = 2S_{21}^\text{H}S_{31}^\text{H}S_{11}^{\text{R}}=T^\text{H}S_{11}^{\text{R}},
\end{equation}
\end{subequations}
where $S_{11}^{\text{R}}$ is the reflection transfer function of the RT phaser given in~\eqref{EQ:RESP_CPLRES_S11}. The quantities $R^{\text{H}}$ and $T^{\text{H}}$ are the reflection and transmission responses of the two-port (1-4) open-ended hybrid network of~\figref{FIG:HYB_2PORTS_OPEN} .

There is a variety of structures that may be used to realize the combined phaser of \figref{FIG:TT_RT_UNIT}. In the next section, we shall consider  an implementation whose RT phaser units are formed by coupled-line systems, which will allow interesting comparison with the TT phaser. In Sec.~\ref{SEC:HYBRID}, the hybrid coupler will be formed using also a coupled-line system for high bandwidth.

\section{Analysis of RT Phaser Unit \\ Realized with Coupled Lines}\label{SEC:ANALYSIS}

\begin{figure}[h!t]
   \centering
   \psfragfig*[width=1\linewidth, trim={0in 0in -0.35in 0in}]{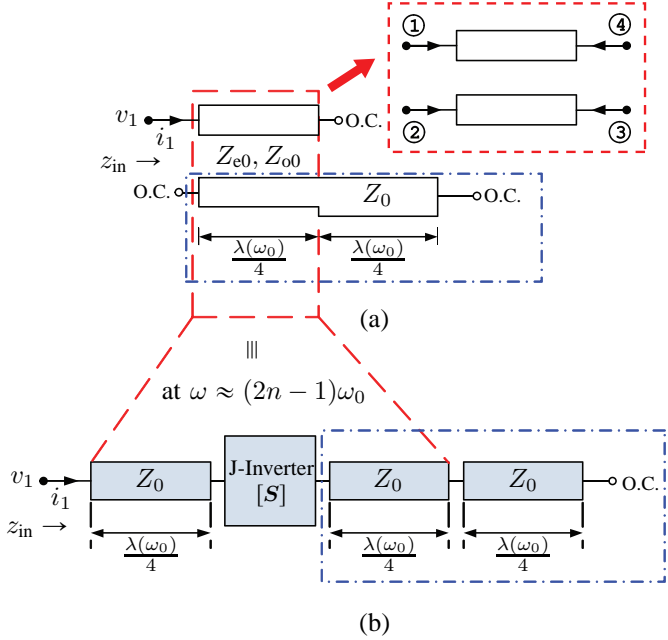}{
    \psfrag{A}[r][r][1]{$v_1$}
    \psfrag{X}[c][c][1]{$i_1$}
    \psfrag{B}[c][c][1]{$Z_0$}
    \psfrag{D}[c][c][1]{$Z_{\text{e0}},Z_{\text{o0}}$}
    \psfrag{b}[c][c][1]{$\equiv$}
    \psfrag{K}[l][l][1]{at $\omega \approx (2n-1)\omega_0$ }
    \psfrag{E}[Tc][Bc][1]{$\frac{\lambda(\omega_0)}{4}$}
    \psfrag{F}[c][c][1]{$z_{\text{in}}\rightarrow$}
    \psfrag{f}[Bc][Tc][1]{$\frac{\lambda(\omega_0)}{4}$}
    \psfrag{a}[c][r][0.83]{$\begin{array}{c}\text{J-Inverter}\\\boldsymbol{[S]}\end{array}$ }
    \psfrag{O}[l][l][0.75]{O.C.}
    \psfrag{g}[c][c][1]{(a)}
    \psfrag{h}[c][c][1]{(b)}
    \psfrag{J}[c][c][0.85]{J-Inverter}
  }
   \caption{Coupled-line implementation of the RT phaser shown in \figref{FIG:IDEAL_UNITS}(b). (a)~Physical configuration. (b)~Equivalent circuit with J-inverter.}
   \label{FIG:RT_CPLIN}
\end{figure}

Figure~\ref{FIG:RT_CPLIN} shows the coupled-line implementation of the RT phaser units in~\figref{FIG:IDEAL_UNITS}. Figure~\ref{FIG:RT_CPLIN}(a) shows the physical coupled-line coupler structure while Fig.~\ref{FIG:RT_CPLIN}(b) shows its inverter equivalent circuit, where a constant delay quarter-wave line section is inserted at the input of the system for the sake of equivalence~\cite{BK:1980_Matthaei}, without any consequence on the group delay swing. The coupled-line coupler section performs the admittance inverter (coupling) function while the bottom half-wavelength stepped-impedance line performs the resonator function.

The equivalence shown in \figref{FIG:RT_CPLIN} holds in the vicinity of odd multiples of the first resonance, $\omega_0$, and can be mathematically expressed as~\cite{BK:2011_Pozar}
\begin{subequations}\label{EQ:RELATION_CPL_JINV1}
\begin{equation}\label{EQ:RELATION_CPL_JINV1_ZE}
  z_{0e} = \dfrac{Z_{0e}}{Z_0}= 1+JZ_0+\left(JZ_0\right)^2=1+\dfrac{1}{\sqrt{\rho_{\text{R}}}}+\dfrac{1}{\rho_{\text{R}}},
\end{equation}
\begin{equation}\label{EQ:RELATION_CPL_JINV1_ZO}
z_{0o}= \dfrac{Z_{0o}}{Z_0} = 1-JZ_0+\left(JZ_0\right)^2=1-\dfrac{1}{\sqrt{\rho_{\text{R}}}}+\dfrac{1}{\rho_{\text{R}}},
\end{equation}
\end{subequations}
where $Z_{0e}$ and $Z_{0o}$ are the even and odd impedances of the coupled-line system in \figref{FIG:RT_CPLIN}(a), respectively, while \eqref{EQ:J_fc_rhoR} was used to obtained the last expressions.

The goal is now to find the input impedance $z_{\text{in}}$, and hence the reflection coefficient of the coupled-line RT phaser to find its group delay characteristics. First, the four-port normalized impedance matrix for the coupled-line coupler shown in \figref{FIG:RT_CPLIN} is expressed in terms of the even and odd impedances as~\cite{BK:2011_Pozar}
\begin{subequations}\label{EQ:zij}
\begin{equation}
  z_{11} = z_{22} = z_{33} = z_{44} = \dfrac{-j}{2}(z_{e0}+z_{o0})\cot\theta,
\end{equation}
\begin{equation}
  z_{12} = z_{21} = z_{34} = z_{43} = \dfrac{-j}{2}(z_{e0}-z_{o0})\cot\theta,
\end{equation}
\begin{equation}
  z_{13} = z_{31} = z_{24} = z_{42} = \dfrac{-j}{2}(z_{e0}-z_{o0})\csc\theta,
\end{equation}
\begin{equation}
  z_{14} = z_{41} = z_{23} = z_{32} = \dfrac{-j}{2}(z_{e0}+z_{o0})\csc\theta.
\end{equation}
\end{subequations}
Next, then open-end boundary conditions at port 2 and 4 shown in \figref{FIG:RT_CPLIN}(a) are applied to reduce the four-port impedance matrix to the two-port matrix
\begin{equation}\label{EQ:Z_MATRIX}
\boldsymbol{[z]}=
  \left[
  \begin{array}{cc}
    z_{11} & z_{13}\\[0.1in]
    z_{31} & z_{33}
  \end{array}
  \right].
\end{equation}
Finally, the boundary condition at port~3, which is the normalized input impedance of the open-end quarter-wavelength transmission line in
\figref{FIG:RT_CPLIN}(a) and reads
\begin{equation}\label{EQ:ZL_AT_PORT3}
  z_{\text{L, }3} = -j\cot\theta=\dfrac{v_3}{i_3},
\end{equation}
The system formed by the $(v_1,v_3)-(i_1,i_3)$ matrix equation associated with \eqref{EQ:Z_MATRIX} plus \eqref{EQ:ZL_AT_PORT3} represents a system of three equations in the variables $v_1,i_1,v_3,i_3$. Eliminating $v_3$ and $i_3$, provides the input impedance for the system in \figref{FIG:RT_CPLIN}(a),
\begin{equation}\label{EQ:Zin_RT}
  z_{\text{in}}=\frac{v_1}{i_1}=z_{11} - \dfrac{z_{13}^2}{z_{11}+z_{\text{L, }3}}.
\end{equation}
The corresponding reflection coefficient is then found as follows. The impedance matrix terms in \eqref{EQ:zij} are written in terms of $\rho_\text{R}$ from the sum and difference of $z_{0e}$ and $z_{0o}$ in \eqref{EQ:RELATION_CPL_JINV1}, and the resulting expressions are inserted in \eqref{EQ:Zin_RT}, which yields
\begin{equation}\label{EQ:RESP_CPLRES_IMP_S11}
\begin{split}
  S_{11\text{, cpl}}^{\text{R}} &= \dfrac{z_{\text{in}}-1}{z_{\text{in}}+1}\\
        & = \dfrac{\cot\theta\left(1+\dfrac{1}{\rho_{\text{R}}}-\dfrac{1}{2\rho_{\text{R}}+1}\sec^2\theta\right)-j}
              {\cot\theta\left(1+\dfrac{1}{\rho_{\text{R}}}-\dfrac{1}{2\rho_{\text{R}}+1}\sec^2\theta\right)+j}.
\end{split}
\end{equation}
where $\theta=\pi\omega/(2\omega_0)$, and where the subscript `cpl' identifies the coupled-line implementation. It is straightforwardly verified in \eqref{EQ:RESP_CPLRES_IMP_S11} that the RT coupled-line phaser is also an all-pass phaser.

The corresponding normalized group delay response can be obtained by taking derivative of $\angle{S_{11\text{, cpl}}^{\text{R}}}(\omega)$ from \eqref{EQ:RESP_CPLRES_IMP_S11}, which yields
\begin{equation}\label{EQ:RESP_CPLRES_IMP_GD}
\begin{split}
  \overline{\tau}_{\text{cpl}}^{\text{R}}
  &= -\dfrac{1}{T_0}\dfrac{\partial\angle{S_{11\text{, cpl}}^{\text{R}}}}{\partial\omega}\\
  &=\dfrac{\csc^2\theta\left(1+\dfrac{1}{\rho_{\text{R}}}-\dfrac{1}{2\rho_{\text{R}}+1}\sec^2\theta\cos2\theta\right)}
          {2+2\cot^2\theta\left(1+\dfrac{1}{\rho_{\text{R}}}-\dfrac{1}{2\rho_{\text{R}}+1}\sec^2\theta\right)^2}.
\end{split}
\end{equation}

Since this RT phaser, represented in \figref{FIG:RT_CPLIN}(a), and the TT (C-section) phaser shown in \figref{FIG:IDEAL_UNITS}(a) are both implemented using coupled lines, their group delay responses can be compared in terms of same parameter, namely the backward coupling factor $k$. However, one must first establish the relationship between $k$ and $\rho_{\text{R}}$ for the coupled-line RT phaser. This relation may be found by inserting~\eqref{EQ:RELATION_CPL_JINV1} into the well-known relation $k={(z_{e0}-z_{o0})}/{(z_{e0}+z_{o0})}$, which yields
\begin{subequations}\label{EQ:REL_K_RHO}
\begin{equation}\label{EQ:REL_K_OF_RHO}
  k=\dfrac{\sqrt{\rho_{\text{R}}}}{\rho_{\text{R}}+1},
\end{equation}
or, solving for $\rho_\text{R}$,
\begin{equation}\label{EQ:REL_RHO_OF_K}
  \rho_{\text{R}} = \dfrac{\left(1+\sqrt{1-4k^2}\right)^2}{4k^2}.
\end{equation}
\end{subequations}

\begin{figure}[h!t]
  \centering
   \psfragfig*[width=1.0\linewidth, trim={0in 0in 0in 0in}]{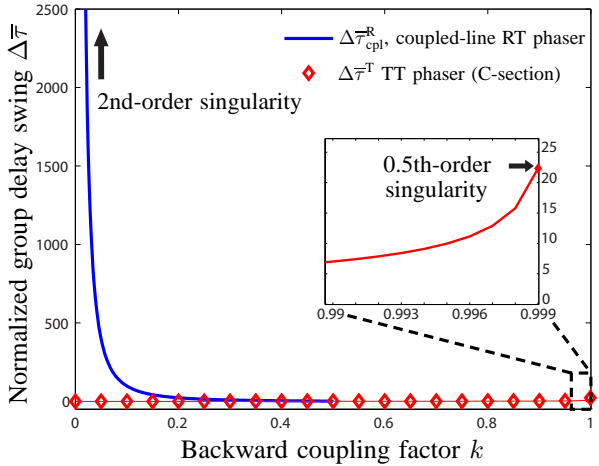}{
    \psfrag{x}[Tc][Bc][1]{Backward coupling factor $k$}
    \psfrag{y}[c][c][1]{Normalized group delay swing $\Delta\overline{\tau}$ }
    \psfrag{a}[Tl][Bl][0.9]{2nd-order singularity}
    \psfrag{b}[l][l][0.9]{0.5th-order}
    \psfrag{c}[l][l][0.9]{singularity}
    \psfrag{C}[l][l][0.75]{$\Delta\overline{\tau}^{\text{R}}_{\text{cpl}}$, coupled-line RT phaser}
    \psfrag{T}[l][l][0.75]{$\Delta\overline{\tau}^{\text{T}}$ TT phaser (C-section)}
  }
  \caption{Comparison of normalized group delay swing $\Delta\overline{\tau}$ for coupled-line RT phaser as shown in \figref{FIG:RT_CPLIN}(a) and TT (C-section) phaser as shown in \figref{FIG:IDEAL_UNITS}(a) with respect to backward coupling factor $k$.}
  \label{FIG:RESP_TT_CLRT_GDSWG}
\end{figure}

The group delay swing for RT coupled-line phaser can now be expressed as a function of $k$. We have
\begin{subequations}\label{EQ:RESP_CPLRES_IMP_GD_EXTREM}
\begin{equation}\label{EQ:RESP_CPLRES_IMP_GD_MAX}
  \overline{\tau}^{\text{R}}_{\text{cpl, max}}
  =\left.\overline{\tau}^{\text{R}}_{\text{cpl}}\right|_{\omega=(2n-1)\omega_0}
  =\rho_{\text{R}}+\frac{1}{2},
\end{equation}
\begin{equation}\label{EQ:RESP_CPLRES_IMP_GD_MIN}
 \overline{\tau}^{\text{R}}_{\text{cpl, min}}
  =\left.\overline{\tau}^{\text{R}}_{\text{cpl}}\right|_{\omega=2n\omega_0}
  =\frac{1}{2}-\dfrac{\rho_{\text{R}}+1}{2\left(2\rho_{\text{R}}^2+2\rho_{\text{R}}+1\right)},
\end{equation}
\end{subequations}
so that
\begin{equation}\label{EQ:RESP_CPLRES_IMP_GDSWG}
\begin{split}
\Delta\overline{\tau}^{\text{R}}_{\text{cpl}}
&=\overline{\tau}^{\text{R}}_{\text{cpl, max}}-\overline{\tau}^{\text{R}}_{\text{cpl, min}}\\
&=\rho_{\text{R}}+\dfrac{\rho_{\text{R}}+1}{2\left(2\rho_{\text{R}}^2+2\rho_{\text{R}}+1\right)}\\
&=\dfrac{\left(1+\sqrt{1-4k^2}\right)^2}{4k^2}+B\left[\rho_{\text{R}}(k)\right],
\end{split}
\end{equation}
where $B\left[\rho_{\text{R}}(k)\right]$ is a bounded function of $\rho_\text{R}$ varying between $0$ ($\rho_\text{R}\rightarrow\infty$ or $k=0$) and 0.2 ($\rho_\text{R}=1$ or $k=1/2$). Therefore, the dominant contribution to the group delay swing in the RT coupled-line phaser is the first term in the last expression of \eqref{EQ:RESP_CPLRES_IMP_GDSWG} near $k\rightarrow 0$, since this term has second-order singularity at $k=0$. This singularity is stronger than the half-order $k=1$ singularity for the TT phaser response in \eqref{EQ:RESP_CSEC_GDSWG}. Figure~\ref{FIG:RESP_TT_CLRT_GDSWG} compares the functions \eqref{EQ:RESP_CPLRES_IMP_GDSWG} and \eqref{EQ:RESP_CSEC_GDSWG} to illustrate this fact. Since the group delay swing variation of the RT phaser near $k=0$ is much sharper than that of the TT phaser near $k=1$, it might be expected that the former will exhibit higher sensitivity to tolerances than the latter. This point will be covered in Sec.~\ref{SEC:SENSITIVITY}.

\begin{figure}[h!t]
  \centering
  \subfigure[]{\label{FIG:GD_CPLRES_IMP_K04}
   \psfragfig*[width=1.0\linewidth, trim={0in 0in 0in 0in}]{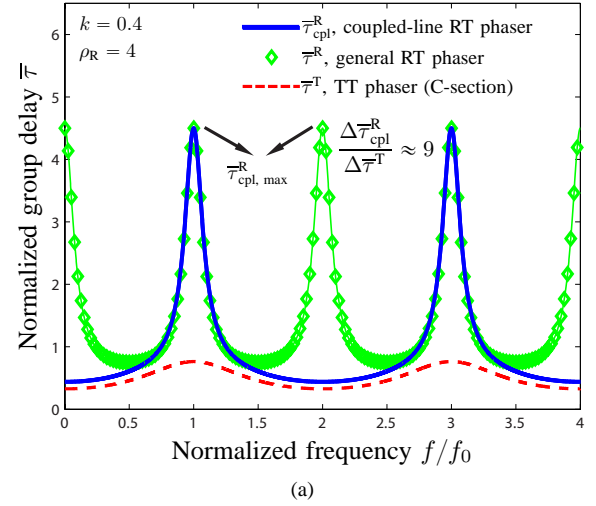}{
    \psfrag{x}[Tc][Bc][1]{Normalized frequency $f/f_0$}
    \psfrag{y}[Bc][Tc][1]{Normalized group delay $\overline{\tau}$}
    \psfrag{C}[l][l][0.75]{$\overline{\tau}^{\text{R}}_{\text{cpl}}$, coupled-line RT phaser}
    \psfrag{G}[l][l][0.75]{$\overline{\tau}^{\text{R}}$, general RT phaser}
    \psfrag{T}[l][l][0.75]{$\overline{\tau}^{\text{T}}$, TT phaser (C-section)}
    \psfrag{k}[l][l][0.75]{$k=0.4$}
    \psfrag{R}[c][c][0.85]{$\dfrac{\Delta\overline{\tau}_\text{cpl}^\text{R}}{\Delta\overline{\tau}^\text{T}}\approx9$}
    \psfrag{r}[l][l][0.75]{$\rho_{\text{R}}=4$}
    \psfrag{P}[c][c][0.75]{$\overline{\tau}^{\text{R}}_{\text{cpl, max}}$}
  }
  }
  \subfigure[]{\label{FIG:GD_CPLRES_IMP_K02}
  \psfragfig*[width=1.0\linewidth, trim={0in 0in 0in 0in}]{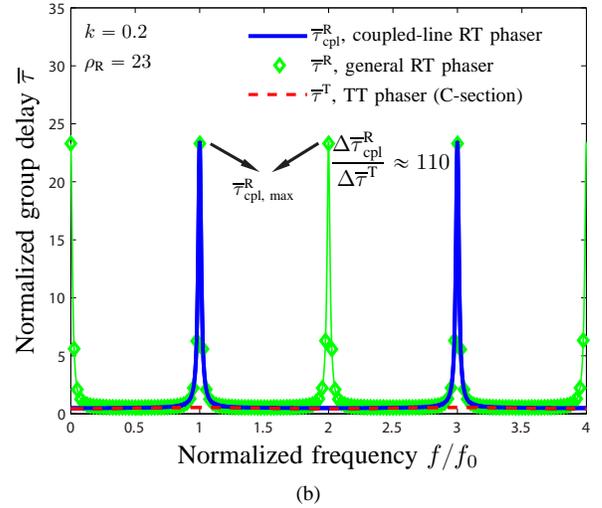}{
    \psfrag{x}[Tc][Bc][1]{Normalized frequency $f/f_0$}
    \psfrag{y}[Bc][Tc][1]{Normalized group delay $\overline{\tau}$}
    \psfrag{C}[l][l][0.75]{$\overline{\tau}^{\text{R}}_{\text{cpl}}$, coupled-line RT phaser}
    \psfrag{G}[l][l][0.75]{$\overline{\tau}^{\text{R}}$, general RT phaser}
    \psfrag{T}[l][l][0.75]{$\overline{\tau}^{\text{T}}$, TT phaser (C-section)}
    \psfrag{k}[l][l][0.75]{$k=0.2$}
        \psfrag{R}[c][c][0.85]{$\dfrac{\Delta\overline{\tau}_\text{cpl}^\text{R}}{\Delta\overline{\tau}^\text{T}}\approx110$}
    \psfrag{r}[l][l][0.75]{$\rho_{\text{R}}=23$}
    \psfrag{P}[c][c][0.75]{$\overline{\tau}^{\text{R}}_{\text{cpl, max}}$}
  }
  }
  \caption{Normalized group delay responses $\overline{\tau}$ versus the normalized frequency $f/f_0$ for the coupled-line RT phaser of \figref{FIG:RT_CPLIN}(a), the general RT phaser of \figref{FIG:IDEAL_UNITS}(b), and the TT phaser of \figref{FIG:IDEAL_UNITS}(a), for (a) $k=0.4$, corresponding to $\rho_{\text{R}} = 4$ according to \eqref{EQ:REL_RHO_OF_K}, and (b) $k=0.2$, corresponding to $\rho_{\text{R}} = 23$.}
  \label{FIG:GD_CPLRES_IMP}
\end{figure}

At this point, the group delay responses of the general RT [\figref{FIG:IDEAL_UNITS}(b)], coupled-line RT [\figref{FIG:RT_CPLIN}(a)] and TT phasers [\figref{FIG:IDEAL_UNITS}(a)] can be compared for the same $k$. This is done in \figref{FIG:GD_CPLRES_IMP}, which plots  $\overline{\tau}_{\text{cpl}}^{\text{R}}(f)$ using \eqref{EQ:RESP_CPLRES_IMP_GD} with \eqref{EQ:REL_RHO_OF_K}, $\overline{\tau}^{\text{R}}(f)$ using \eqref{EQ:RESP_CPLRES_GD} and $\overline{\tau}^{\text{T}}(f)$ using \eqref{EQ:RESP_CSEC_GD} for two different values of $k$. Figure~\ref{FIG:GD_CPLRES_IMP} shows that for a given $k$, the RT phasers provide much higher group swings than the RT phaser, as predicted in Sec.~\ref{SEC:COMPARISON_TT_RT}. Note that the period of the coupled-line RT phaser is twice the period of the general RT phaser because the coupled-line coupler used in the former case supports odd harmonics only. Finally, ~\figref{FIG:GD_CPLRES_IMP} also shows how the group delay swing frequency response of the RT phasers decreases as $k$ increases, consistently with \figref{FIG:RESP_TT_CLRT_GDSWG}. The reason for this effect is that increasing $k$ increases the splitting between the coupled (even and ode) modes and hence the resonance bandwidth, which reduces the group delay swing, since the area under the group delay curve of a phaser is conserved \cite{JOUR:2010_TMTT_Gupta}. Thus, to increase the group delay swing, and hence the R-ASP resolution, in coupled-line RT phaser, on just needs to decrease the coupling factor $k$, i.e. increase the separation between the coupled-lines, which incidentally eases fabrication, whereas the opposite holds for a TT coupler.

\section{Choice of Hybrid Coupler for \\ the Proposed Combined Phaser}\label{SEC:HYBRID}

An ideal lossless combined phaser would exhibit perfect matching and perfect transmission, i.e. $S_{11}^{\text{C}}=0$ in~\eqref{EQ:RESP_TT_RT_UNIT_S11} and $S_{21}^{\text{C}}=1$ in~\eqref{EQ:RESP_TT_RT_UNIT_S21}, respectively. According to~\eqref{EQ:RESP_TT_RT_UNIT}, this would require the hybrid coupler to exhibit a scattering response such that $|R^{\text{H}}|=|(S_{21}^\text{H})^2+(S_{31}^\text{H})^2|=0$ and \mbox{$|T^{\text{H}}|=|2S_{21}^\text{H}S_{31}^\text{H}|=1$}. Since this requirement is practically necessarily limited in bandwidth, any hybrid coupler implementation will somewhat affect the magnitude and group delay responses of the overall combined phaser.

\begin{figure}[h!t]
  \centering
   \psfragfig*[width=0.95\linewidth, trim={0in 0in 0in 0in}]{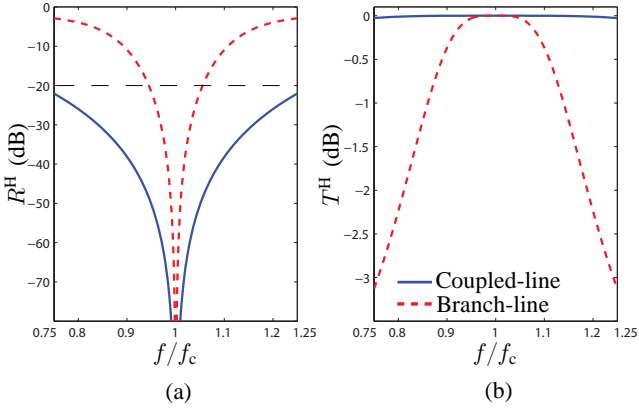}{
    \psfrag{x}[Tc][Bc][0.9]{$f/f_{\text{c}}$}
    \psfrag{y}[c][c][0.9]{$R^{\text{H}}$ (dB)}
    \psfrag{z}[c][c][0.9]{$T^{\text{H}}$ (dB)}
    \psfrag{C}[l][l][0.9]{Coupled-line}
    \psfrag{B}[l][l][0.9]{Branch-line}
    \psfrag{a}[c][c][0.9]{(a)}
    \psfrag{b}[c][c][0.9]{(b)}
  }
  \caption{Analytical responses in \figref{FIG:HYB_2PORTS_OPEN} for a branch-line quadrature hybrid coupler (red dashed curves) and for a backward coupled-line hybrid coupler (blue solid lines). (a)~Reflection coefficient $|R^{\text{H}}(f)|$ in~\eqref{EQ:RESP_TT_RT_UNIT_S11}. (b)~Transmission coefficient $|T^{\text{H}}(f)|$ in~\eqref{EQ:RESP_TT_RT_UNIT_S21}.}
  \label{FIG:RESP_BLHYB_2PORTS_OPEN}
\end{figure}

The dashed curves in~\figref{FIG:RESP_BLHYB_2PORTS_OPEN} show the $|R^{\text{H}}(f)|$ and $|T^{\text{H}}(f)|$ responses of a branch-line hybrid coupler~\cite{BK:2011_Pozar}, which was used in \cite{CONF:2014_IMS_Zou} as a starting point towards the design of a broadband combined phaser. As expected, the $|R^{\text{H}}|=0$-$|T^{\text{H}}|=1$ bandwidth is quite limited, $10\%$ at $20$~dB for $|R^{\text{H}}|$, due to the discrete nature (4~connection points) of its coupling/interference mechanism.

\begin{figure}[h!t]
  \centering
   \psfragfig*[width=1.0\linewidth, trim={0in 0in 0in 0in}]{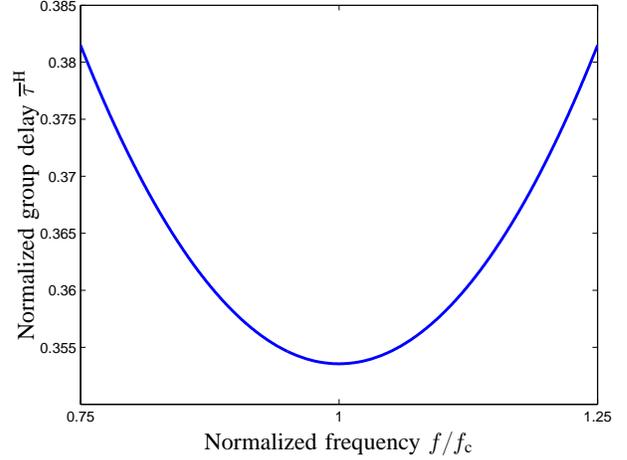}{
    \psfrag{x}[Tc][Bc][0.9]{Normalized frequency $f/f_{\text{c}}$}
    \psfrag{y}[c][c][0.9]{Normalized group delay $\overline{\tau}^{\text{H}}$}
  }
  \caption{Normalized (with respect to the center frequency) group delay response of the coupled-line hybrid coupler.}
  \label{FIG:RESP_GD_CLHYB_2PORTS_OPEN}
\end{figure}

Much higher bandwidth can be obtained with a coupled-line coupler~\cite{BK:2011_Pozar}, due to the distributed nature of its coupling/interference mechanism. This is shown in \figref{FIG:RESP_BLHYB_2PORTS_OPEN}, where the coupled-line coupler $|R^{\text{H}}|$ $20$-dB bandwidth reaches $50\%$. Even larger bandwidth could naturally be obtained by resorting to multi-section coupled-line configurations.
The effect of the hybrid on the group delay response of the combined phaser may be quantified by adding the the derivative of $\angle{T^{\text{H}}}(f)$ to the group delay of the RT phaser units. The result is plotted in \figref{FIG:RESP_GD_CLHYB_2PORTS_OPEN}, which shows that over the $|R^\text{H}(f)|$ 20~dB bandwidth the group delay variation is within $3\%$ of $T_{\text{c}}$, where $T_{\text{c}}$ is the period at the center (quarter-wavelength) frequency of the coupler. Since practical group delay swings exceed this variation by at least 2 order of magnitude, the group delay introduced by the hybrid can be considered constant over the operating bandwidth of the combined phaser. The overall group delay of the combined phaser is thus
\begin{equation}\label{EQ:GD_TT_RT}
 {\tau}^{\text{C}}=T_0\overline{\tau}^{\text{R}}_{\text{cpl}}+T_{\text{c}}\overline{\tau}^{\text{H}}\approx{T_0}\overline{\tau}^{\text{R}}_{\text{cpl}}+\text{const.},
\end{equation}
where $\overline{\tau}^{\text{R}}_{\text{cpl}}$ is given in \eqref{EQ:RESP_CPLRES_IMP_GD}. Equation~\eqref{EQ:GD_TT_RT} means that the coupled-line hybrid is essentially transparent to the RT phasers connected at its coupled and through ports, since the constant term disappears in the group delay swing, which is the quantity of interest in R-ASP.

In the forthcoming design, the coupled-line hybrid coupler is implemented in slot coupled-line technology~\cite{BK:2007_Mongia} is used, and the corresponding results will be given in Sec.~\ref{SEC:IMP}.

\section{Synthesis of the Combined Phaser \\ for Prescribed Group Delay} \label{SEC:SYNTHESIS}

We set as a goal to synthesize a phaser with a linear group delay (quadratic phase) between angular frequencies $\omega_{\text{L}}$ and $\omega_{\text{U}}$, i.e.
\begin{equation}\label{EQ:Desired_GD}
    \tau_{\text{p}}(\omega) = \sigma(\omega-\omega_{\text{L}})+\text{const.},
\end{equation}

\noindent where $\sigma$ is the (constant) group delay slope. This is the response required in a real-time spectrum analyser, for instance~\cite{Jour:2013_MwMag_Caloz}.

\begin{figure}[h!t]
  \centering
   \psfragfig*[width=1.0\linewidth]{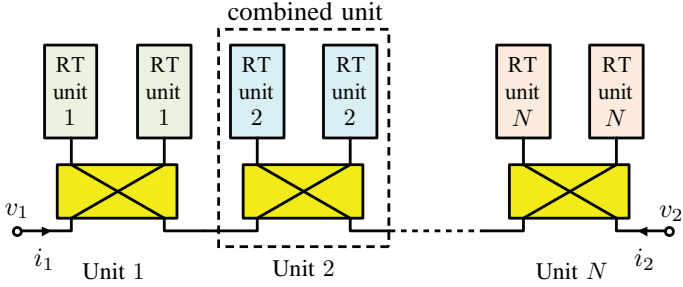}{
   \psfrag{e}[Bc][Tc][1]{$v_1$}
   \psfrag{f}[Tc][Bc][1]{$i_1$}
   \psfrag{g}[Bc][Tc][1]{$v_2$}
   \psfrag{h}[Tc][Bc][1]{$i_2$}
   \psfrag{N}[c][c][0.9]{$N$}
   \psfrag{1}[c][c][0.9]{$1$}
   \psfrag{2}[c][c][0.9]{$2$}
   \psfrag{G}[c][c][1]{combined unit}
   \psfrag{a}[c][c][0.9]{Unit $1$}
   \psfrag{b}[c][c][0.9]{Unit $2$}
   \psfrag{c}[c][c][0.9]{Unit $N$}
  }
  \caption{Complete phaser formed by cascading $N$ combined phasers, each of which consists of a hybrid coupler and a pair of identical coupled-line RT units of the type shown in \figref{FIG:RT_CPLIN}(a).}
  \label{FIG:combined_Phaser}
\end{figure}

\begin{figure}[h!t]
  \centering
   \psfragfig*[width=1.0\linewidth, trim={0in 0in 0in 0in}]{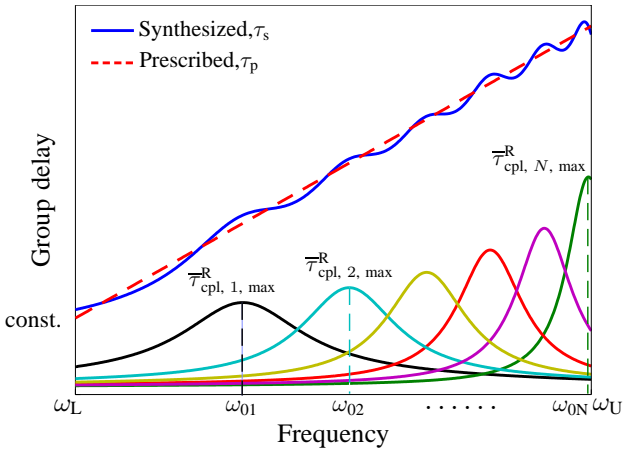}{
    \psfrag{x}[Tc][Bc][1]{Frequency}
    \psfrag{y}[Bc][Tc][1]{Group delay}
    \psfrag{O}[l][l][0.85]{Synthesized,$\tau_{\text{s}}$}
    \psfrag{P}[l][l][0.85]{Prescribed,$\tau_{\text{p}}$ }
    \psfrag{l}[c][c][1]{$\omega_{\text{L}}$ }
    \psfrag{h}[l][l][1]{$\omega_{\text{U}}$ }
    \psfrag{d}[c][c][1.1]{$\ldots\ldots$ }
    \psfrag{C}[r][r][0.9]{const. }
    \psfrag{1}[c][c][0.9]{$\omega_{\text{01}}$}
    \psfrag{2}[c][c][0.9]{$\omega_{\text{02}}$}
    \psfrag{4}[r][r][0.9]{$\omega_{\text{0N}}$}
    \psfrag{s}[Bc][tc][0.85]{$\overline{\tau}^{\text{R}}_{\text{cpl, }1\text{, max}}$}
    \psfrag{t}[c][c][0.85]{$\overline{\tau}^{\text{R}}_{\text{cpl, }2\text{, max}}$}
    \psfrag{v}[Br][tr][0.85]{$\overline{\tau}^{\text{R}}_{\text{cpl, }N\text{, max}}$}
  }
  \caption{Synthesized group delay response for each section (corresponding to each peak) and the summation of them, $\tau_{\text{s}}$, to fit the prescribed one $\tau_{\text{p}}$. }
  \label{FIG:SYNTHESIS}
\end{figure}

A phaser with quasi\footnote{``Quasi'' means here ``within the physical limits of causality.''} arbitrary group delay response may be realized by cascading a sufficient number, $N$, of combined phaser sections, as shown in \figref{FIG:combined_Phaser}, which contribute group delay maximum at different frequencies to form the overall delay response, as illustrated in \figref{FIG:SYNTHESIS}. Note that the RT phaser units in each combined phaser sections must be identical to avoid imbalance, which would result in spurious reflection. Using identical RT phasers in each section theoretically ensures zero reflection ($S_{11}^\text{C}=0$) and complete transmission ($|S_{21}^\text{C}|=1$), since the RT phaser units are all-pass networks ($|S_{11}^\text{R}|=1$), if the hybrid coupler is ideal ($R^\text{H}=0$ and $|T^\text{H}|=1$), as seen in~\eqref{EQ:RESP_TT_RT_UNIT}. Although a practical implementation necessarily deviates from these ideal conditions, the departure from the ideal magnitude responses will be relatively minor, and only the group delay has to be considered in the synthesis.

The overall group delay response of the phaser is
\begin{equation}\label{EQ:Synthesized_GD}
    \tau_{\text{s}}(\omega) = \sum_{i=1}^{N}T_{0i}\overline{\tau}^{\text{R}}_{\text{cpl,}i}\left[\rho_{\text{R}i}(k_i), \theta_{i}(\omega_{0i})\right]+\text{const.},
\end{equation}
where $T_{\text{0}i}$ is the period corresponding to resonant frequency $\omega_{0i}$ ($T_{\text{0}i}=2\pi/\omega_{0i}$) of the $i^{\text{th}}$ unit, $\rho_{\text{R}i}$ is SWR of the J-inverter of the $i^{\text{th}}$ unit, and $\theta_{i} = \pi\omega/(2\omega_{\text{0}i})$ is the electrical length of the coupler of  the $i^{\text{th}}$ unit. The additional group delay introduced by the hybrids in \eqref{EQ:Synthesized_GD} are absorbed in the constant term since they are nondispersive (see Sec.~\ref{SEC:HYBRID}).

The synthesis consists in minimizing the error between the prescribed and synthesized group delay responses,
\begin{equation}\label{EQ:ERROR}
    e(\omega) = \left|\dfrac{\tau_{\text{s}}(\omega)-\tau_{\text{p}}(\omega)}{\tau_{\text{p}}(\omega)}\right|^2,
\end{equation}
over the frequency band of interest in order to obtain the parameter sets $\{\omega_{\text{0}i}\}$ and $\{k_i\}$, \mbox{$i=1,2,\ldots,N$}, where the first set provides the physical lengths of the coupled lines while the second set provides the even/odd impedances of the couplers via \eqref{EQ:RELATION_CPL_JINV1} with \eqref{EQ:REL_RHO_OF_K}. The open-end capacitance of each transmission line caused by fringe electric field should be taken into account in determining the physical length of the line~\cite{BK:2011_Hong}. The hybrids transforming the pairs of coupled-line RT phaser units into the combined phaser are identically tuned (quarter-wavelength) at the operation band center frequency, $\omega_{\text{c}}=(\omega_{\text{L}}+\omega_{\text{U}})/2$. Moreover, $\omega_{\text{L}}$ and $\omega_{\text{U}}$ are chosen such that the fractional bandwidth, $(\omega_{\text{U}}-\omega_{\text{L}})/\omega_{\text{c}}$, is smaller than 50$\%$ to fit within the bandwidth of the hybrid couplers \figref{FIG:RESP_BLHYB_2PORTS_OPEN}.

Note that~\eqref{EQ:ERROR} is completely known analytically. Therefore, the synthesis is extremely fast, requiring only minor full-wave tunig.

\section{Implementation demonstration}\label{SEC:IMP}

\begin{figure}[h!t]
  \centering
   \psfragfig*[width=1.0\linewidth, trim={0in 0in 0in 0in}]{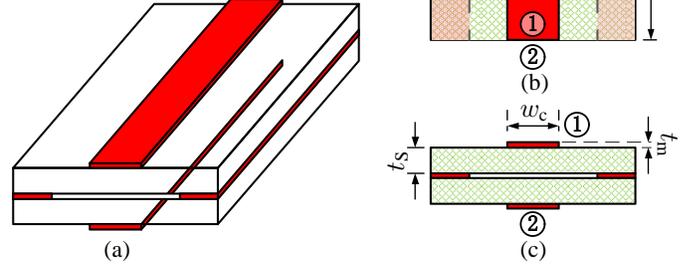}{
   \psfrag{E}[l][l][0.9]{Copper}
   \psfrag{F}[l][l][0.9]{Substrate Rogers RO3210}
    \psfrag{x}[Tc][Bc][0.9]{(a)}
    \psfrag{y}[Bc][Tc][0.9]{(b)}
    \psfrag{z}[Tc][Bc][0.9]{(c)}
    \psfrag{a}[Bc][Tc][1]{$w_{\text{c}}$}
    \psfrag{b}[c][c][1]{$w_{\text{s}}$}
    \psfrag{c}[Bc][Tc][1]{$\ell_{\text{c}}=\ell_{\text{s}}$}
    \psfrag{d}[c][c][1]{$t_{\text{S}}$}
    \psfrag{e}[Bc][Tc][1]{$t_{\text{m}}$}
    \psfrag{T}[Bl][Tl][1]{\textbf{Substrate parameters}}
    \psfrag{D}[Tl][Tl][1]{\textbf{Dimensions (UNIT: MIL)}}
    \psfrag{S}[Tl][Bl][1]{$\begin{array}{ll}
    \varepsilon_{\text{r}}=10.8 & t_{\text{S}}=25\text{ mil} \\
    \tan\delta = 0.003 &  t_{\text{m}} = 17.5 \text{ $\mu$m}
    \end{array}$}
    \psfrag{L}[Tl][Bl][1]{$\begin{array}{ll}
    w_{\text{c}}=82.64 & w_{\text{s}}=289.24\\
    \ell_{\text{c}}=\ell_{\text{s}}=310
    \end{array}$}
  }
  \caption{Slot-coupled coupled-line hybrid coupler and corresponding dimensions (subscripts c and s denote conductor and slot parameters, respectively). Ports 1 and 4 are on one side, and port 2 and 3 are on the other side of the coupling slot. (a) Perspective view. (b) Top view. (c) Transverse cross-sectional view.}
  \label{FIG:SLOT_CPLHYB}
\end{figure}

We synthesize here an $N=6$ combined unit phasers prototype (\figref{FIG:combined_Phaser}) with a linear group delay of swing $\Delta\tau=5$~ns over the frequency range $[f_L,f_U]=[3,4.8]$~GHz. In order to easily achieve hybrid (3-dB) coupling, we use slot-coupled coupled-line couplers~\cite{BK:2007_Mongia} to combine the RT phaser units. Related dimensions and material parameters are shown in \figref{FIG:SLOT_CPLHYB}.

\begin{figure}[h!t]
  \centering
   \psfragfig*[width=0.9\linewidth, trim={0in 0in 0in 0in}]{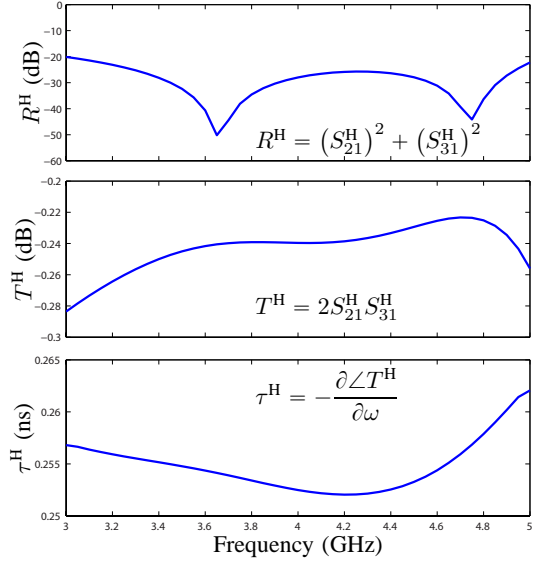}{
   \psfrag{h}[Tc][Bc][0.9]{Frequency (GHz)}
   \psfrag{x}[c][c][0.9]{$R^{\text{H}}$ (dB)}
   \psfrag{y}[c][c][0.9]{$T^{\text{H}}$ (dB)}
   \psfrag{z}[c][c][0.9]{$\tau^{\text{H}}$ (ns)}
   \psfrag{R}[l][l][0.9]{$R^{\text{H}}=\left(S_{21}^\text{H}\right)^2+\left(S_{31}^\text{H}\right)^2$}
   \psfrag{T}[l][l][0.9]{$T^{\text{H}}=2S_{21}^\text{H}S_{31}^\text{H}$}
   \psfrag{G}[Tl][Bl][0.9]{$\tau^{\text{H}}=-\dfrac{\partial\angle{T^{\text{H}}}}{\partial\omega}$}
  }
  \caption{Full-wave simulated responses for the slot-coupled coupled-line hybrid coupler shown in \figref{FIG:SLOT_CPLHYB} in the configuration of \figref{FIG:HYB_2PORTS_OPEN}.}
  \label{FIG:RESP_SHYB_2PORTS_OPEN}
\end{figure}

The full-wave responses $R^{\text{H}}(f)$ and $T^{\text{H}}(f)$ of the slot-coupled coupled-line hybrid coupler are plotted in \figref{FIG:RESP_SHYB_2PORTS_OPEN} (Experimental responses are not accessible since the coupler is integrated in the overall phaser system). The reflection coefficient $R^{\text{H}}$ is lower than -20~dB over the bandwidth 3 -- 5 GHz, which is greater than the specified bandwidth of interests (3 -- 4.8 GHz). The insertion loss introduced by the slot-coupled coupled-line hybrid coupler is less than 0.3~dB. Its group delay variation is $\Delta\tau^\text{H}=0.009$ ns, corresponding to a relative variation of $\Delta\tau^\text{H}/\tau^\text{H}_\text{avg}=3.5\%$ with respect to the average group delay value $\tau_\text{avg}^\text{H}=0.255$~ns. The group delay can thus be considered constant in the operation bandwidth, as expected from Sec.~\ref{SEC:HYBRID}.

\begin{figure}[h!t]
  \centering
   \psfragfig*[width=0.9\linewidth, trim={0in 0in 0in 0in}]{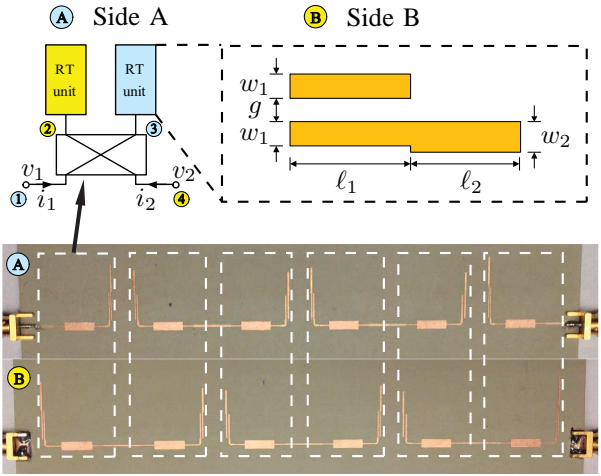}{
   \psfrag{w}[cc][cc][1]{$w_{1}$}
   \psfrag{G}[cc][cc][1]{$g$}
   \psfrag{q}[Tc][Bc][1]{$\ell_{1}$}
   \psfrag{p}[cc][cc][1]{$w_{2}$}
   \psfrag{a}[Tc][Bc][1]{$\ell_{2}$}
   \psfrag{x}[l][l][1]{Side A}
   \psfrag{y}[l][l][1]{Side B}
   \psfrag{e}[l][l][1]{$v_1$}
   \psfrag{f}[l][l][1]{$i_1$}
   \psfrag{g}[l][l][1]{$v_2$}
   \psfrag{h}[r][l][1]{$i_2$}
  }
  \caption{Layout of the 6-unit phaser prototype corresponding to the configuration of \figref{FIG:combined_Phaser}, with geometrical parameters listed in Tab.~1.}
  \label{FIG:Layout}
\end{figure}

The complete phaser, corresponding to the configuration of \figref{FIG:combined_Phaser} with $N=6$ combined phaser units, is shown in \figref{FIG:Layout}. For each combined phaser, one coupled-line RT phaser is connected to the coupled port (port 2) of the slot hybrid on one side and the other one is connected to the through port (port 4) of the slot hybrid on the other side. The parameters $k_i$ and $f_{0i}$ and corresponding geometrical parameters obtained by synthesis for each combined phaser are listed in \mbox{Tab.~1}.
\begin{table}[h!t]
    \caption{Electrical [$k$, $f_0$ (GHz)] and geometrical (mil) parameters for the six-unit combined phaser shown in \figref{FIG:Layout}.}
    \centering
    \begin{tabular}{c|cccccc}\label{TAB:Params}
      Unit $i$         & 1         & 2         & 3         & 4         & 5         & 6\\[1pt]
      \hline\\[-8pt]
       $k$          & 0.3969    & 0.3563    & 0.3426    & 0.3241    & 0.2897    & 0.2404    \\[1pt]
       $f_0$        & 3.6370    & 4.0033    & 4.2751    & 4.4715    & 4.6439    & 4.8055    \\[1pt]
       \hline
       $w_{1}$      & 11.07   & 13.08   & 13.69   & 14.47   & 15.77   & 17.38   \\[1pt]
       $g$          & 6.45    & 7.79    & 8.34    & 9.16    & 10.96   & 14.38   \\[1pt]
       $\ell_{1}$   & 299.72  & 274.99  & 255.49  & 246.00  & 235.29  & 226.00  \\[1pt]
       $w_{2}$      & 21.3    & 21.3    & 21.3    & 21.3    & 21.3    & 21.3      \\[1pt]
       $\ell_{2}$   & 293.19  & 267.63  & 249.45  & 237.81  & 228.98  & 220.63  \\[1pt]
    \end{tabular}
 \end{table}

\begin{figure}[h!t]
  \centering
   \psfragfig*[width=1.0\linewidth, trim={0in 0in 0in 0in}]{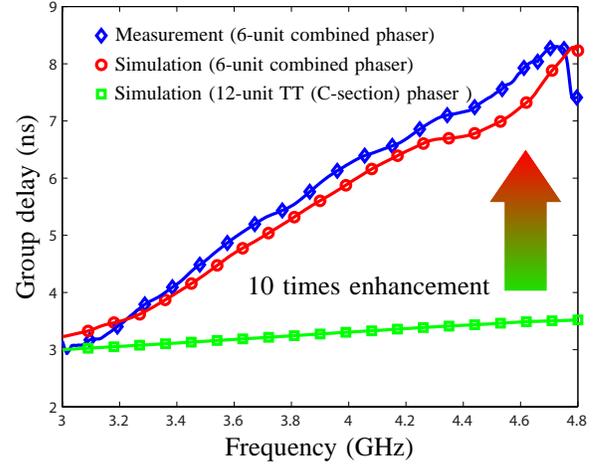}{
    \psfrag{x}[Tc][Bc][1]{Frequency (GHz)}
    \psfrag{y}[Bc][Tc][1]{Group delay (ns)}
    \psfrag{M}[l][l][0.75]{Measurement (6-unit combined phaser)}
    \psfrag{S}[l][l][0.75]{Simulation (6-unit combined phaser)}
    \psfrag{E}[tr][br][1]{10 times enhancement }
    \psfrag{C}[l][l][0.75]{Simulation (12-unit TT (C-section) phaser ) }
  }
  \caption{Experimental (blue diamond) and numerical (red circle) results of group delay response for the fabricated six-unit combined phaser as shown in \figref{FIG:Layout} in comparison with numerical result (green square) for a 12-unit TT phaser (C-section), 10 times enhancement in group delay swing is observed.}
  \label{FIG:GD_TM}
\end{figure}

Figure~\ref{FIG:GD_TM} shows the group delay response for the fabricated six-unit combined phaser. In order to demonstrate the sought R-ASP resolution enhancement, the figure also shows the result for a 12-unit TT (C-section) phaser, which represents 6 pairs of RT phasers for comparison with the proposed phaser. It appears that the proposed phaser provides a tenfold group delay swing enhancement compared its TT counterpart.

\begin{figure}[h!t]
  \centering
   \subfigure[]{\label{FIG:MAG_S21}
   \psfragfig*[width=1.0\linewidth, trim={0in 0in 0in 0in}]{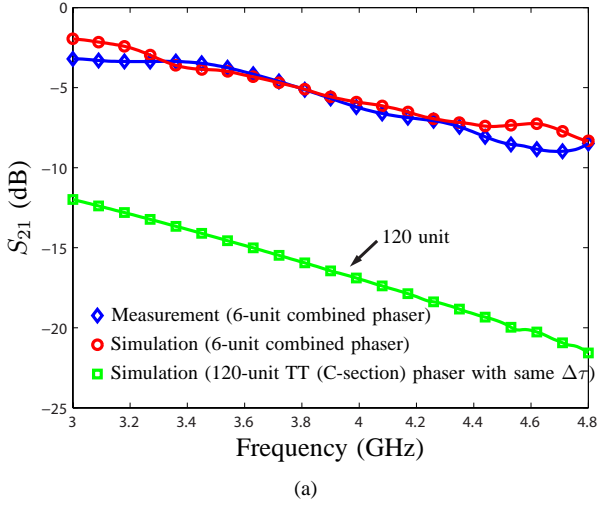}{
    \psfrag{x}[Tc][Bc][1]{Frequency (GHz)}
    \psfrag{y}[Bc][Tc][1]{$S_{21}$ (dB)}
    \psfrag{M}[l][l][0.75]{Measurement (6-unit combined phaser)}
    \psfrag{S}[l][l][0.75]{Simulation (6-unit combined phaser)}
    \psfrag{B}[l][l][0.75]{120 unit }
    \psfrag{C}[l][l][0.75]{Simulation (120-unit TT (C-section) phaser with same $\Delta\tau$) }
  }}
  \subfigure[]{\label{FIG:MAG_S11}
   \psfragfig*[width=1.0\linewidth, trim={0in 0in 0in 0in}]{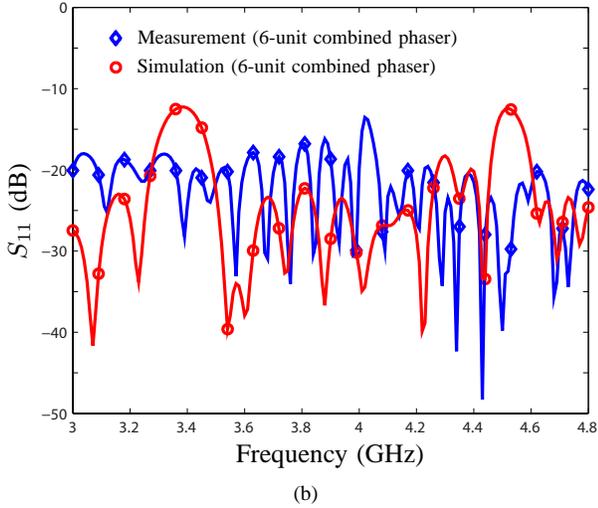}{
    \psfrag{x}[Tc][Bc][1]{Frequency (GHz)}
    \psfrag{y}[Bc][Tc][1]{$S_{11}$ (dB)}
    \psfrag{M}[l][l][0.75]{Measurement (6-unit combined phaser)}
    \psfrag{S}[l][l][0.75]{Simulation (6-unit combined phaser)}
  }}
  \caption{Magnitude response for the fabricated six-unit combined phaser shown in \figref{FIG:Layout}. (a)~$S_{21}$, comparison with the result obtained by simulation for a 120-unit TT (C-section) phaser (green squares) with same group delay swing (5~ns). (b)~Reflection coefficient, $S_{11}$.}
  \label{FIG:MAG}
\end{figure}

In terms of the magnitude response, good agreement between the experimental and numerical results for $S_{21}$ are observed in \figref{FIG:MAG_S21}, where a the response of a 120-unit TT phaser providing the same group delay swing is also shown. It appears, due to its high number of required units, the TT phaser has an insertion loss that is almost three times larger than that of the proposed phaser. Figure~\ref{FIG:MAG_S11} shows that the reflection coefficient is below -15 dB over almost the entire operation band.

\section{Sensitivity analysis}\label{SEC:SENSITIVITY}

Figure~\ref{FIG:RESP_TT_CLRT_GDSWG} showed that, in contrast to the TT phaser, the RT coupled-line phaser exhibits a sharp group delay swing enhancement as $k$ decreases towards zero, because of the second-order singularity at $k=0$ in the group delay swing, given by \eqref{EQ:RESP_CPLRES_IMP_GDSWG}. However, such a sharp variation could imply high sensitivity to the tolerances of the line widths $w$ and gaps $g$ (Fig.~\ref{FIG:Layout}), which would naturally be practically unfavorable. The fabrication tolerance in the process used for the fabrication of the prototype is of $\delta=\pm{3}$.

\begin{figure}[h!t]
  \centering
   \psfragfig*[width=0.9\linewidth, trim={0in 0in 0in 0in}]{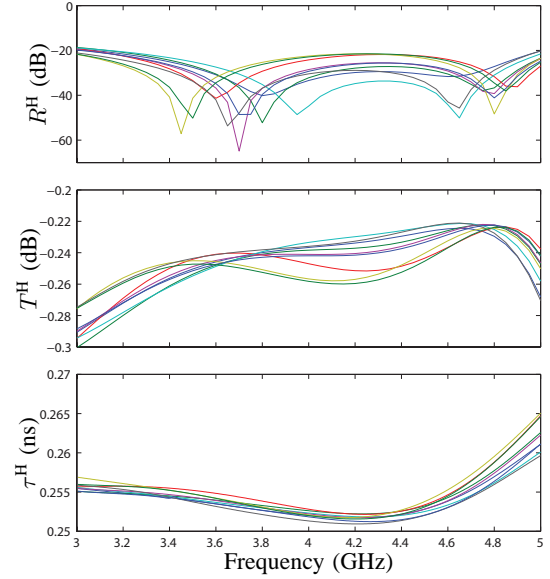}{
   \psfrag{h}[Tc][Bc][0.9]{Frequency (GHz)}
   \psfrag{x}[c][c][0.9]{$R^{\text{H}}$ (dB)}
   \psfrag{y}[c][c][0.9]{$T^{\text{H}}$ (dB)}
   \psfrag{z}[c][c][0.9]{$\tau^{\text{H}}$ (ns)}
  }
  \caption{Sensitivity to tolerance ($\delta=\pm3\text{ mil}$) analysis for the slot-coupled coupled-line hybrid coupler configured as in \figref{FIG:HYB_2PORTS_OPEN} considering various $\delta$ changes of the nominal values of $w_{\text{c}}$ and $w_{\text{s}}$ in \figref{FIG:SLOT_CPLHYB}.}
  \label{FIG:RESP_SHYB_2PORTS_OPEN_TOLERANCE}
\end{figure}

We first analyze the tolerance variations of the slot-coupled coupled-line hybrid coupler with coupled and through ports open as in \figref{FIG:HYB_2PORTS_OPEN}. Full-wave simulation results for the sensitivity to tolerance are shown in \figref{FIG:RESP_SHYB_2PORTS_OPEN_TOLERANCE}. It is seen that the coefficient $R^\text{H}$ remains below -20 dB for all typical parameter deviations within the fabrication tolerances. The transmission coefficient $T^{\text{H}}$ remains greater than -0.3 dB. And the relative variation of $\tau^{\text{H}}$ is within $4.5\%$. Thus, coupler is not significantly sensitive to tolerances. Tolerance sensitivity comparison with the TT phaser can then be restricted to the RT coupled-line phaser.

The tolerance sensitivity relations for the TT and RT coupled-line phasers are as follows
\begin{subequations}\label{EQ:SENS_TT}
\begin{equation}\label{EQ:SENS_TTa}
\dfrac{\partial\Delta\overline{\tau}^{\text{T}}}{\partial{w}}=
\dfrac{d\Delta\overline{\tau}^{\text{T}}}{d{k}}\dfrac{\partial{k}}{\partial{w}},\quad
\dfrac{\partial\Delta\overline{\tau}^{\text{T}}}{\partial{g}}=
\dfrac{d\Delta\overline{\tau}^{\text{T}}}{d{k}}\dfrac{\partial{k}}{\partial{g}},
\end{equation}
where
\begin{equation}\label{EQ:SENS_TT_DT_OVER_DK}
 \dfrac{d\Delta\overline{\tau}^{\text{T}}}{d{k}}=\dfrac{1}{\left(1-k^2\right)^{\frac{3}{2}}},
\end{equation}
\end{subequations}
and
\begin{subequations}\label{EQ:SENS_RT}
\begin{equation}\label{EQ:SENS_RTa}
\dfrac{\partial\Delta\overline{\tau}^{\text{R}}_{\text{cpl}}}{\partial{w}}=
\dfrac{d\Delta\overline{\tau}^{\text{R}}_{\text{cpl}}}{d{k}}\dfrac{\partial{k}}{\partial{w}},\quad
\dfrac{\partial\Delta\overline{\tau}^{\text{R}}_{\text{cpl}}}{\partial{g}}=
\dfrac{d\Delta\overline{\tau}^{\text{R}}_{\text{cpl}}}{d{k}}\dfrac{\partial{k}}{\partial{g}},
\end{equation}
where
\begin{equation}\label{EQ:SENS_RT_DT_OVER_DK}
 \dfrac{d\Delta\overline{\tau}^{\text{R}}_{\text{cpl}}}{d{k}}=
  \dfrac{d\Delta\overline{\tau}^{\text{R}}_{\text{cpl}}}{d{\rho_{\text{R}}}}\dfrac{d\rho_{\text{R}}}{dk},
\end{equation}
with
\begin{equation}\label{EQ:SENS_RT_DT_OVER_RHO}
  \dfrac{d\Delta\overline{\tau}^{\text{R}}_{\text{cpl}}}{d{\rho_{\text{R}}}} =
  1-\dfrac{1}{2\left(2\rho_{\text{R}}^2+2\rho_{\text{R}}+1\right)}-\dfrac{\rho_{\text{R}}}{\left(2\rho_{\text{R}}^2+2\rho_{\text{R}}+1\right)^2},
\end{equation}
\begin{equation}\label{EQ:SENS_RT_RHO_OVER_DK}
  \dfrac{d\rho_{\text{R}}}{dk}=-\dfrac{\left(1+\sqrt{1-4k^2}\right)^2}{2k^3\sqrt{1-4k^2}}.
\end{equation}
\end{subequations}

The sensitivity formulas in \eqref{EQ:SENS_TTa} and \eqref{EQ:SENS_RTa} are constituted of two multiplicative terms. One is the sensitivity of the group delay swing with respect to the coupling factor. The other one is the sensitivity of $k$ with respect to the parameters $w$ or $g$. The two terms will be next analyzed for $k<0.5$, which represents a coupling factor range accessible to planar fabrication.

\begin{figure}[h!t]
  \centering
   \psfragfig*[width=1\linewidth, trim={0in 0in 0in 0in}]{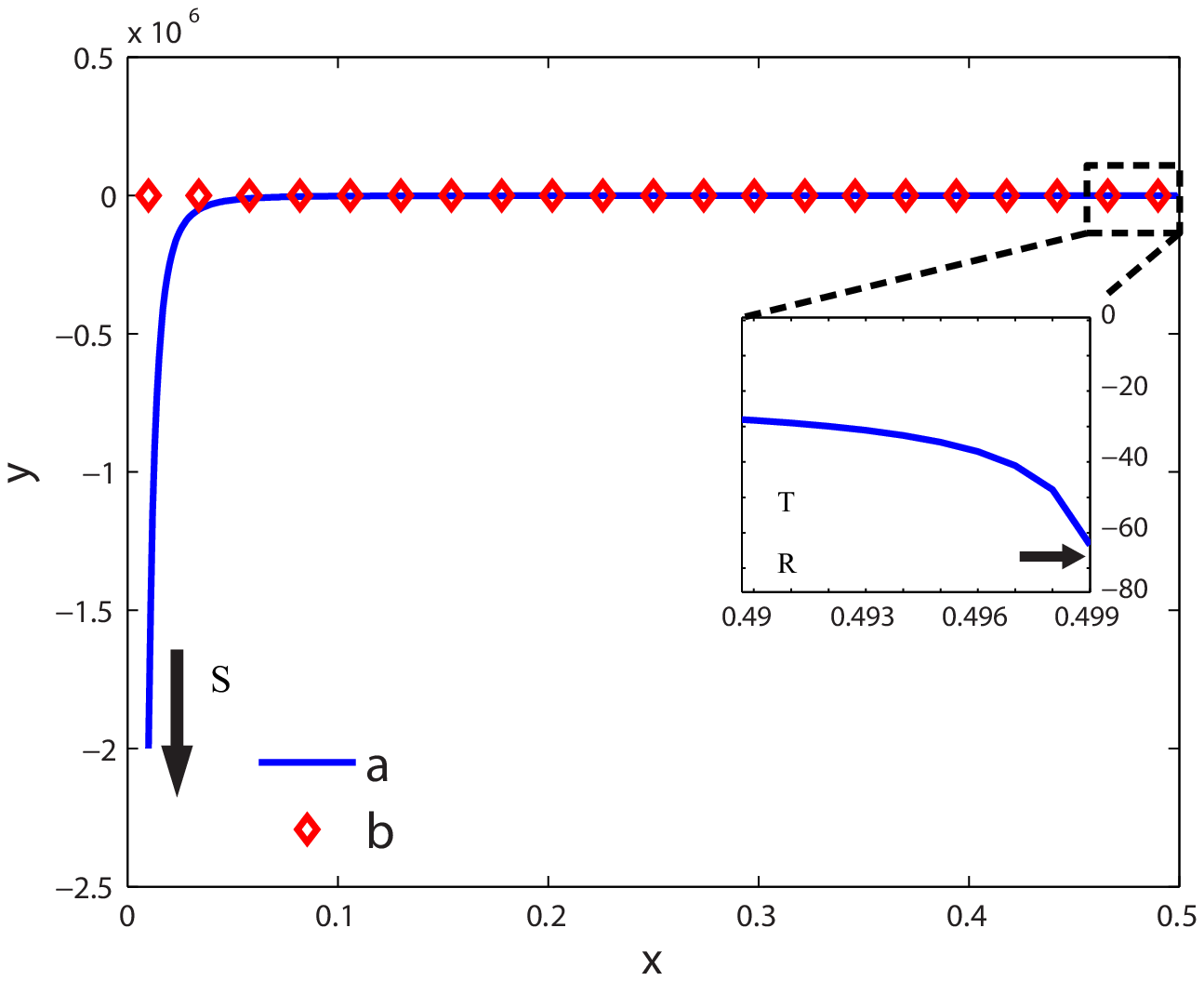}{
   \psfrag{x}[Tc][Bc][0.9]{Backward coupling factor $k$}
   \psfrag{y}[c][c][0.9]{${d\Delta\overline{\tau}}/{dk}$}
   \psfrag{S}[l][l][0.9]{3rd-order singularity}
   \psfrag{T}[l][l][0.9]{1.5th-order}
   \psfrag{R}[l][l][0.9]{singularity}
   \psfrag{a}[l][l][0.75]{${d\Delta\overline{\tau}^{\text{R}}_{\text{cpl}}}/{d{k}}$ (29b) for coupled-line RT phaser}
   \psfrag{b}[l][l][0.75]{${d\Delta\overline{\tau}^{\text{T}}}/{d{k}}$ (28b) for TT phaser (C-section)}
  }
  \caption{Comparison of ${d\Delta\overline{\tau}^{\text{R}}_{\text{cpl}}}/{d{k}}$ [Eq.~\eqref{EQ:SENS_RT_DT_OVER_DK}] and  ${d\Delta\overline{\tau}^{\text{T}}}/{d{k}}$ [Eq.~\eqref{EQ:SENS_TT_DT_OVER_DK}].}
  \label{FIG:DT_OVER_DK}
\end{figure}

Equation~\eqref{EQ:SENS_TT_DT_OVER_DK} shows that the sensitivity of $\Delta\overline{\tau}^{\text{T}}$ with respect to $k$ includes a 1.5-order singularity at $k=1$, which is outside the practical region $k<0.5$. For the RT coupled-line phaser, the sensitivity of $\Delta\overline{\tau}^{\text{R}}_{\text{cpl}}$ with respect to $k$ given in \eqref{EQ:SENS_RT_DT_OVER_DK} is the product of \eqref{EQ:SENS_RT_DT_OVER_RHO} and \eqref{EQ:SENS_RT_RHO_OVER_DK}. Equation~\eqref{EQ:SENS_RT_RHO_OVER_DK} has 3rd-order singularity at $k=0$ and 0.5-order singularity at $k=0.5$. Since $\rho_{\text{R}}$ has a 2nd-order singularity at $k=0$, as shown in \eqref{EQ:REL_RHO_OF_K}, the second and third terms in \eqref{EQ:SENS_RT_DT_OVER_RHO} have 4th-order and 6th-order zeros at $k=0$, respectively. Therefore, the product of \eqref{EQ:SENS_RT_DT_OVER_RHO} and \eqref{EQ:SENS_RT_RHO_OVER_DK} in \eqref{EQ:SENS_RT_DT_OVER_DK} has 1st-order and 3rd-order zeros at $k=0$ for the second and third terms, and 3rd-order singularity at $k=0$ for the first term. In summary, the RT coupled-line phaser sensitivity of $\Delta\overline{\tau}^{\text{R}}_{\text{cpl}}$ with respect to $k$ in \eqref{EQ:SENS_RT_DT_OVER_DK} has a 3rd-order singularity at $k=0$ and 0.5th-order singularity at $k=0.5$, while the TT phaser sensitivity of $\Delta\overline{\tau}^{\text{T}}$ with respect to $k$ in \eqref{EQ:SENS_TT_DT_OVER_DK} has a 1.5th-order singularity at $k=1$, which is outside the practical region. Figure~\ref{FIG:DT_OVER_DK} shows the sensitivity of $\Delta\overline{\tau}$ with respect to $k$ as given in \eqref{EQ:SENS_TT_DT_OVER_DK} and \eqref{EQ:SENS_RT_DT_OVER_DK}, respectively. The high-order singularity at $k=0$ for the RT coupled-line phaser is evident.

\begin{figure}[h!t]
  \centering
   \psfragfig*[width=1\linewidth, trim={0in 0in 0in 0in}]{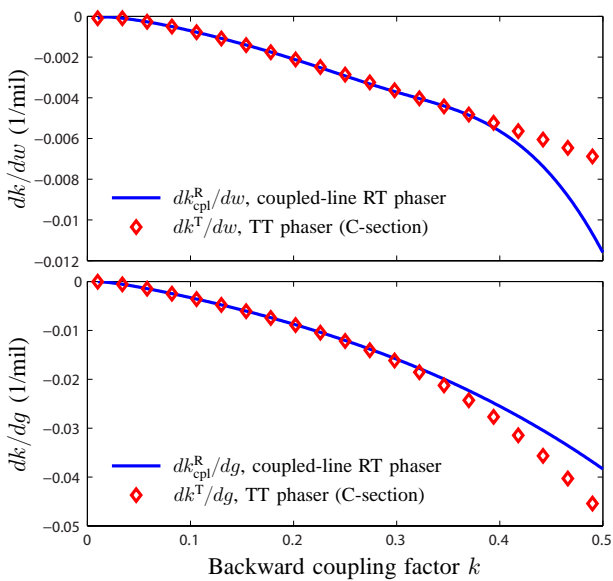}{
   \psfrag{x}[Tc][Bc][0.9]{Backward coupling factor $k$}
   \psfrag{y}[c][c][0.9]{${dk}/{dw}$ (1/mil)}
   \psfrag{z}[c][c][0.9]{${dk}/{dg}$ (1/mil)}
   \psfrag{a}[l][l][0.75]{ $dk^{\text{R}}_{\text{cpl}}/{d{g}}$, coupled-line RT phaser }
   \psfrag{b}[l][l][0.75]{ $dk^{\text{T}}/{d{g}}$, TT phaser (C-section) }
   \psfrag{c}[l][l][0.75]{ $dk^{\text{R}}_{\text{cpl}}/{d{w}}$, coupled-line RT phaser }
   \psfrag{d}[l][l][0.75]{ $dk^{\text{T}}/{d{w}}$, TT phaser (C-section) }
  }
  \caption{Sensitivity of $k$ with respect to the parameters $w$ and $g$.}
  \label{FIG:DK_OVER_DW(DG)}
\end{figure}

The second multiplicative terms in \eqref{EQ:SENS_TTa} and \eqref{EQ:SENS_RTa} are the sensitivities of $k$ with respect to geometrical parameters. Figure~\ref{FIG:DK_OVER_DW(DG)} shows $\partial{k^{\text{R}}_{\text{cpl}}}/\partial{w}$, $\partial{k^{\text{T}}}/\partial{w}$, $\partial{k^{\text{R}}_{\text{cpl}}}/\partial{g}$ and $\partial{k^{\text{T}}}/\partial{g}$ versus $k$, obtained numerically. Two points should be mentioned. First, the sensitivity of $k$ with respect to $w$ and $g$ has opposite-slope variation compared to $d\Delta\overline{\tau}/dk$, and $k=0$ represents now a zero instead of a singularity, which is obvious since $k=0$ corresponds to the case of uncoupled lines. Secondly, the sensitivity of $k$ to $g$ variations is about five times higher than that to variation in $w$.

\begin{figure}[h!t]
  \centering
   \psfragfig*[width=1\linewidth, trim={0in 0in 0in 0in}]{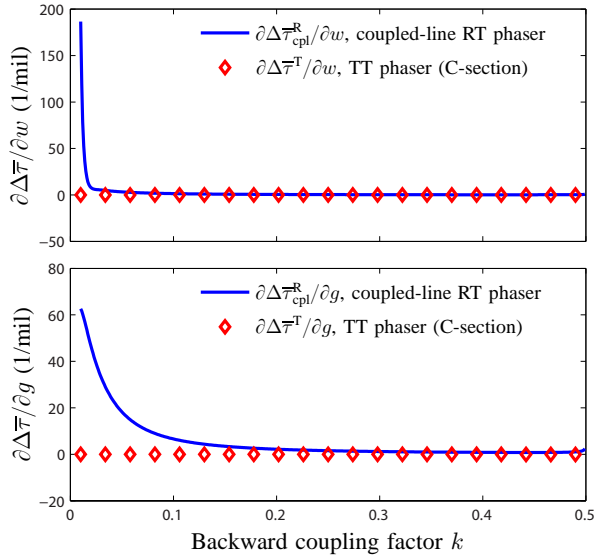}{
   \psfrag{x}[Tc][Bc][0.9]{Backward coupling factor $k$}
   \psfrag{y}[c][c][0.9]{${\partial\Delta\overline{\tau}}/{\partial{g}}$ (1/mil)}
   \psfrag{z}[c][c][0.9]{${\partial\Delta\overline{\tau}}/{\partial{w}}$ (1/mil)}
   \psfrag{a}[l][l][0.75]{${\partial\Delta\overline{\tau}^{\text{R}}_{\text{cpl}}}/{\partial{w}}$, coupled-line RT phaser }
   \psfrag{b}[l][l][0.75]{${\partial\Delta\overline{\tau}^{\text{T}}}/{\partial{w}}$, TT phaser (C-section) }
   \psfrag{c}[l][l][0.75]{${\partial\Delta\overline{\tau}^{\text{R}}_{\text{cpl}}}/{\partial{g}}$, coupled-line RT phaser }
   \psfrag{d}[l][l][0.75]{${\partial\Delta\overline{\tau}^{\text{T}}}/{\partial{g}}$, TT phaser (C-section) }
  }
  \caption{Partial derivative of group delay swing $\Delta\overline{\tau}$ with respect to dimensions $w$ and $g$, which are conductor width and gap for coupled lines in RT unit shown in \figref{FIG:RT_CPLIN}(a) and \figref{FIG:IDEAL_UNITS}(a) respectively. }
  \label{FIG:DT_OVER_DW(DG)}
\end{figure}

Finally, the overall sensitivity of $\Delta\overline{\tau}$ with respect to $w$ and $g$ is plotted in \figref{FIG:DT_OVER_DW(DG)}. It may be seen that the high sensitivity singularity of $\Delta\overline{\tau}^{\text{R}}_{\text{cpl}}$ near $k=0$ is partly compensated by the
sensitivity zero of $k$ with respect to $w$ and $g$, which is naturally a favorable result.

\begin{figure}[h!t]
  \centering
   \psfragfig*[width=1\linewidth, trim={0in 0in 0in 0in}]{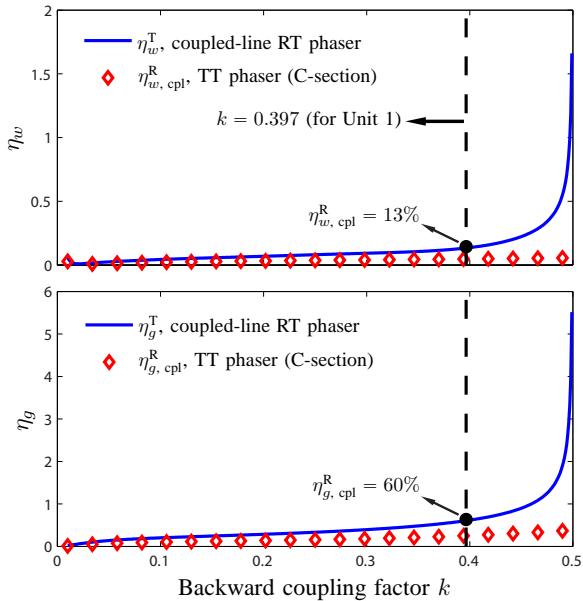}{
   \psfrag{x}[Tc][Bc][0.9]{Backward coupling factor $k$}
   \psfrag{y}[c][c][0.9]{$\eta_{w}$}
   \psfrag{z}[c][c][0.9]{$\eta_{g}$}
   \psfrag{a}[l][l][0.75]{$\eta_{w}^{\text{T}}$, coupled-line RT phaser }
   \psfrag{b}[l][l][0.75]{$\eta_{w\text{, cpl}}^{\text{R}}$, TT phaser (C-section) }
   \psfrag{c}[l][l][0.75]{$\eta_{g}^{\text{T}}$, coupled-line RT phaser }
   \psfrag{d}[l][l][0.75]{$\eta_{g\text{, cpl}}^{\text{R}}$, TT phaser (C-section) }
   \psfrag{k}[r][r][0.75]{$k=0.397$ (for Unit 1)}
   \psfrag{e}[r][r][0.75]{$\eta_{w\text{, cpl}}^{\text{R}}=13\%$}
   \psfrag{f}[r][r][0.75]{$\eta_{g\text{, cpl}}^{\text{R}}=60\%$}
  }
  \caption{Relative deviation of group delay swing with when fabrication tolerance is \mbox{$\pm3\text{ mil}$}.}
  \label{FIG:REL_DEV_DT}
\end{figure}

More relevant is the relative deviation of the group delay swing with respect to the fabrication tolerance $\delta$,
\begin{subequations}\label{EQ:REL_DEV_DT}
\begin{equation}
  \eta^{\text{T}}_{w}=\dfrac{\left|\dfrac{\partial\Delta\overline{\tau}^{\text{T}}}{\partial{w}}\delta\right|}{\Delta\overline\tau^{\text{T}}},\quad
  \eta^{\text{T}}_{g}=\dfrac{\left|\dfrac{\partial\Delta\overline{\tau}^{\text{T}}}{\partial{g}}\delta\right|}{\Delta\overline\tau^{\text{T}}},
\end{equation}
\begin{equation}
  \eta^{\text{R}}_{w\text{, cpl}}
  =\dfrac{\left|\dfrac{\partial\Delta\overline{\tau}^{\text{R}}_\text{cpl}}{\partial{w}}\delta\right|}{\Delta\overline\tau^{\text{R}}_\text{cpl}},\quad
  \eta^{\text{R}}_{g\text{, cpl}}
  =\dfrac{\left|\dfrac{\partial\Delta\overline{\tau}^{\text{R}}_\text{cpl}}{\partial{g}}\delta\right|}{\Delta\overline\tau^{\text{R}}_\text{cpl}},
\end{equation}
\end{subequations}
These functions are plotted in \figref{FIG:REL_DEV_DT}. Contrary to what might be suggested from \figref{FIG:RESP_TT_CLRT_GDSWG}, the singularity at $k=0$ in \eqref{EQ:SENS_RT} and \eqref{EQ:SENS_RT} is totally canceled out by the singularity in the group delay swing function ($\Delta\overline{\tau}^{\text{T}}$ for TT phasers and $\Delta\overline{\tau}^{\text{R}}_{\text{cpl}}$ for RT phasers), which yields zero sensitivity at $k=0$. Thus, we have simultaneously high group delay swing and low sensitivity in the $k=0$ region, which represents a most favorable situation for practical design.

Note the emergence of a singularity at $k=0.5$ for the coupled-line RT phaser. Considering the highest coupling factor (0.397) among the 6 units in the fabricated prototype, the relative deviation of group delay swing could be as high as $60\%$ when the gap varies to $g_0\pm3$ mil, where $g_0$ is nominal value for $k=0.397$. This issue may be relaxed by  enforcing all the RT coupled-line phaser units to operate in a lower $k$ region.

\section{Conclusion}\label{SEC:CONCLUSION}

A high Radio Analog Signal Processing (R-ASP) resolution transmission-type (TT) phaser based on reflection-type (RT) phaser units has been presented and analyzed. It has been shown, theoretically and experimentally, that RT phasers inherently exhibit higher R-ASP resolution than their TT counterparts because their group delay swing is proportional to the reflection coefficient of the coupling system (admittance inverter) rather than to the coupling coefficient (TT case), where the former can easily by maximized towards unity whereas the latter is practically limited to relatively low values. A detailed sensitivity analysis has revealed that the proposed phaser is simultaneously characterized by high R-ASP resolution and low sensitivity to fabrication tolerance.

Given its favourable characteristics, the proposed phaser might significantly contribute to promote R-ASP from a conceptual level to a commercial reality for tomorrow's radio.

\bibliographystyle{IEEEtran}
\bibliography{IEEEabrv,REF_MTT}

%

\end{document}